\documentclass[twocolumn]{aastex63}

\usepackage{graphicx}   
\usepackage{subfigure}
\usepackage{multirow}
\usepackage{amsmath}
\usepackage{lineno}

\graphicspath{{./}{figures/}}
\shorttitle{Impact of cosmic web filaments on galaxies}
\shortauthors{Yu, Zhu, Yang, Mo, Luan \& Feng.}

\begin{document}
\title{Impact of cosmic web on the properties of galaxies in IllustrisTNG simulations}

\correspondingauthor{Weishan Zhu}
\email{zhuwshan5@mail.sysu.edu.cn}

\author[0009-0001-6368-2833]{Guangyao, Yu}
\affil{School of Physics and Astronomy, Sun Yat-Sen University, Zhuhai campus, No. 2, Daxue Road \\
Zhuhai, Guangdong, 519082, China}
\affil{CSST Science Center for the Guangdong-Hong Kong-Macau Greater Bay Area, Daxue Road 2, 519082, Zhuhai, China}

\author[0000-0002-1189-2855]{Weishan, Zhu}
\affil{School of Physics and Astronomy, Sun Yat-Sen University, Zhuhai campus, No. 2, Daxue Road \\
Zhuhai, Guangdong, 519082, China}
\affil{CSST Science Center for the Guangdong-Hong Kong-Macau Greater Bay Area, Daxue Road 2, 519082, Zhuhai, China}

\author{Qi-Rui, Yang}
\affil{School of Physics and Astronomy, Sun Yat-Sen University, Zhuhai campus, No. 2, Daxue Road \\
Zhuhai, Guangdong, 519082, China}
\affil{CSST Science Center for the Guangdong-Hong Kong-Macau Greater Bay Area, Daxue Road 2, 519082, Zhuhai, China}

\author{Jian-Feng, Mo}
\affil{School of Physics and Astronomy, Sun Yat-Sen University, Zhuhai campus, No. 2, Daxue Road \\
Zhuhai, Guangdong, 519082, China}
\affil{CSST Science Center for the Guangdong-Hong Kong-Macau Greater Bay Area, Daxue Road 2, 519082, Zhuhai, China}

\author[0009-0001-6592-6467]{Tian-cheng Luan}
\affil{School of Physics and Astronomy, Sun Yat-Sen University, Zhuhai campus, No. 2, Daxue Road \\
Zhuhai, Guangdong, 519082, China}
\affil{CSST Science Center for the Guangdong-Hong Kong-Macau Greater Bay Area, Daxue Road 2, 519082, Zhuhai, China}

\author{Long-Long, Feng}
\affil{School of Physics and Astronomy, Sun Yat-Sen University, Zhuhai campus, No. 2, Daxue Road \\
Zhuhai, Guangdong, 519082, China}
\affil{CSST Science Center for the Guangdong-Hong Kong-Macau Greater Bay Area, Daxue Road 2, 519082, Zhuhai, China}



\begin{abstract}
We investigate the influence of the cosmic web on galaxy properties in the IllustrisTNG simulations. To disentangle the effects of galaxy groups and cosmic filaments, we classify the cosmic web environment into four categories: group, group-dominated, filament-dominated, and field. By controlling for stellar mass, we reveal evident differences in specific star formation rates (sSFR), quenched fraction, gas fractions, local density, and stellar ages among central galaxies in different cosmic web environments, particularly for lower-mass galaxies. However, these differences largely diminish when the effect of local overdensity is further accounted for, indicating its dominant role. Additionally, we observe distinct differences in these properties among satellite galaxies across environments, mainly driven by stellar mass, halo mass, and overdensity. Notably, residual differences between satellites in field and filament-dominated region persist even after controlling for these factors, suggesting a stronger susceptibility of satellite galaxies to filaments compared to centrals. Our findings highlight the importance of differentiating between central and satellite to accurately assess the environmental effects of the cosmic web. Our analysis suggests that the relationship between galaxy properties and their distance from filaments arises from a combination of factors, including stellar and halo mass, groups, overdensity, and the intrinsic influence of the cosmic web. Additionally, we find that the effect of the cosmic web on galaxy properties is reduced at $z=0.5$, compared to $z=0$. Furthermore, central galaxies near thick filaments tend to exhibit slightly to moderately lower sSFR and cold gas fractions compared to those near thin filaments.

\end{abstract}


\section{Introduction} \label{sec:intro}

In the past several decades, a comprehensive theory of galaxy formation and evolution developed within the framework of $\Lambda \rm{CDM}$ cosmology (e.g. \citealt{white1978core,1991ApJ...379...52W,mo2010galaxy}). In this paradigm, galaxies form at the centers of dark matter halos, assembling their gas and stars through accretion and mergers. According to this model, galaxy properties are shaped by both internal and external influences. Internally, galaxy evolution is governed by physical processes such as gas accretion, cooling and heating, star formation, and feedback from massive stars and active galactic nuclei (AGN) (\citealt{white1978core,1991ApJ...379...52W, 2004MNRAS.353..713K,primack2024galaxy}). Externally, the surrounding environment of a galaxy can also play an important role in shaping its properties. The environment of galaxies can be generally classified into three scales: the halo scale (approximately the virial radius), the local environment (within $\sim1-2\, h^{-1}$ Mpc), and the cosmic web environment (beyond $\sim1-2\, h^{-1}$ Mpc). Note that the first two scales may overlap for groups and clusters. 

At the halo scale, it is well established that halo mass and mechanisms such as harassment, strangulation, and ram pressure stripping play a crucial role in galaxy formation and evolution (e.g. \citealt{gunn1972infall,1991ApJ...379...52W, mo2010galaxy,2018ARA&A..56..435W,xie2020influence,wang2023environmental}). At the local environment scale, the halo mass function varies with local overdensity, with massive halos predominantly residing in high-density regions, leading to a strong correlation between galaxy properties and local overdensity (e.g.\citealt{1980ApJ...236..351D,2004MNRAS.353..713K,2010ApJ...721..193P,2018ARA&A..56..435W}). In addition, the interaction and merge rate in the high over-density region would be higher.

The large-scale spatial distribution of galaxies exhibits a distinct cosmic web pattern (e.g.,\citealt{1986ApJ...302L...1D,2003astro.ph..6581C,2014MNRAS.438..177A,2014MNRAS.438.3465T}),which can be classified into nodes (clusters), filaments, sheets (walls), and voids. Theoretical studies suggest that the cosmic web arises from the anisotropic gravitational collapse of matter in the nonlinear regime (\citealt{1970A&A.....5...84Z,1996Natur.380..603B,2008LNP...740..335V}). Its formation and evolution have been extensively studied through cosmological simulations (\citealt{2005MNRAS.359..272C,2007A&A...474..315A,2009MNRAS.396.1815F,2010ApJ...723..364A,2010MNRAS.408.2163A,2012MNRAS.425.2049H,2014MNRAS.441.2923C,2017ApJ...838...21Z,2018MNRAS.473.1195L}). However, whether the cosmic web and, in particular, its filamentary structures, significantly influences galaxy properties remains an open question and a subject of ongoing debate.

Numerous observational studies suggest that galaxies in different cosmic web environments exhibit distinct properties. For example, galaxies in regions of low local density are, on average, bluer, less massive, and have later-type morphologies. They exhibit higher cold gas content, elevated star formation rates (SFRs), slower stellar mass growth histories, lower stellar metallicities, and a more gradual evolutionary pace compared to their counterparts in denser large-scale environments (e.g. \citealt{2004ApJ...617...50R,2007ApJ...658..898P,2012MNRAS.426.3041H,2021ApJ...906...97F,dominguez2023galaxies,zakharova2024virgo}). Moreover, several studies have reported that galaxies more closer to filaments generally tend to have lower SFRs, older ages, redder colors, higher metallicities, and greater masses (\citealt{alpaslan2016galaxy, chen2017detecting, kraljic2018galaxy,2018MNRAS.474.5437L,winkel2021imprint,castignani2022virgo,laigle2025euclid}). In addition, \cite{donnan2022role} finds that galaxies closer to nodes exhibit higher gas-phase metallicities, with a similar but weaker trend observed for filaments.

Several factors contribute to these trends. One major driver is the relationship between galaxy properties and local environmental density (\citealt{1980ApJ...236..351D,2004MNRAS.353..713K}). Another key factor is the mass of the host halo. Recent studies have attempted to isolate the additional influence of the cosmic web on galaxies by disentangling the effects of local overdensity and halo mass. \cite{kuutma2017voids} finds that even after accounting for environmental density and redshift, residual trends remain as galaxies approach filaments: early-type galaxies become more prevalent, color index increase, and sSFRs decline. \cite{kleiner2017evidence} report that for galaxies with stellar masses below $10^{11}\,\mathrm{M_\odot}$, the $\mathrm{H\,\textsc{i}}$ fraction is comparable between those near filaments and those in the field. In contrast, \cite{odekon2018effect} find that, at fixed local density and stellar mass, the $\mathrm{H\,\textsc{i}}$ deficiency of galaxies with stellar masses between $10^{8.5}\,\mathrm{M_\odot}$ and $10^{10.5}\,\mathrm{M_\odot}$ becomes more pronounced as they approach filaments. \cite{hoosain2024effect} suggests that the observed reddening and gas depletion of galaxies near filaments are primarily driven by the increased presence of galaxy groups in these regions, with filaments themselves exerting only a minor effect on the gas content. \cite{o2024effect} reports that differences in star formation activity and morphological fractions between filament and field galaxies disappear once the local galaxy density is controlled for.

To accurately assess the influence of the cosmic web on galaxies, it is essential to disentangle the effects of stellar mass, halo mass, local density, and the impact of galaxy groups and clusters. Additionally, the influence of cosmic web may differ between central and satellite galaxies due to their distinct evolutionary pathways. However, observational studies face several challenges in effectively separating these factors, including determining host halo masses, identifying galaxy groups, and distinguishing satellite galaxies from low-mass centrals. Cosmological hydrodynamical simulations of galaxy formation and evolution provide a powerful tool to address these challenges. In particular, several state-of-the-art simulations, such as Horizon-AGN (\citealt{2014MNRAS.444.1453D}), EAGLE (\citealt{schaye2015eagle}), IllustrisTNG (\citealt{2018MNRAS.473.4077P}) and SIMBA (\citealt{2019MNRAS.486.2827D}) can reproduce many statistical properties of observed galaxy populations. Recent studies have explored the role of the cosmic web, particularly filaments, in shaping galaxy properties using cosmological simulations (\citealt{2019MNRAS.483.3227K,song2021beyond,2022A&A...658A.113M,2022MNRAS.514.2488Z,hasan2023filaments, bulichi2024galaxy, ma2024neutraluniversemachine}). Most of these studies find that galaxies located closer to nodes and filaments tend to be more massive, exhibit lower sSFRs, and have higher quenched fractions, consistent with observational findings. However, only a few works have effectively disentangled the effects of local overdensity and halo mass, suggesting a possible additional influence of the cosmic web environment on galaxy evolution (\citealt{2019MNRAS.483.3227K,song2021beyond,hasan2023filaments,o2024effect}).

In this study, we investigate the impact of the cosmic web on galaxy properties using data from the IllustrisTNG simulations. Our analysis distinguishes between the influences of galaxy groups and filaments while controlling for the effects of halo mass and local environment density. Additionally, we examine central and satellite galaxies separately. Section \ref{sec:method} provides an overview of the IllustrisTNG simulations, details the selected galaxy samples and the properties analyzed, and outlines the methodologies used for filament identification and environmental classification. 
In Section \ref{sec:results}, we first examine the impact of environment on central galaxy properties, including results after correcting for overdensity. We then explore the effects on satellite galaxies, both before and after controlling for halo mass and overdensity. Then we show the results for the whole galaxy sample. Additionally, we investigate the relationship between galaxy properties and their distance from filaments. Section \ref{sec:discussions} extends our analysis to higher redshifts, compares results across different resolutions of the TNG simulations, and examines the distinct effects of thick and thin filaments. Section \ref{sec:conclusions} summarizes our key findings.

Throughout this paper, we adopt the cosmological parameters from Planck 2015: $\Omega_m = 0.3089$,  $\Omega_b = 0.0486$, $\Omega_\lambda = 0.6911$,  $\mathit{H}_\mathrm{0} = 67.74 \, \mathrm{km \, s}^{-1} \mathrm{Mpc}^{-1}$ \citep{ade2016planck}, align with the IllustrisTNG simulations.

\section{Methodology} \label{sec:method}

\subsection{IllustrisTNG Simulations}

The IllustrisTNG simulations (\citealt{nelson2018first, pillepich2018first, springel2018first}) builds upon the original Illustris project \citep{nelsondylan_2015}, utilizing the AREPO moving mesh code \citep{springel2010pur}. These simulations track the evolution of dark matter, gas, stars, and supermassive black holes while incorporating key baryonic processes such as radiative cooling, star formation, chemical enrichment, and stellar and AGN feedback. 

IllustrisTNG consists of three simulation suites: TNG50, TNG100, and TNG300, with volumes of $(51.7\, \mathit{c}\mathrm{Mpc})^3$, $(110.7\, \mathit{c}\mathrm{Mpc})^3$, and $(302.6\, \mathit{c}\mathrm{Mpc})^3$, respectively. The mass of a single dark matter particle is $4.5 \times 10^{5}\, \mathrm{M}_{\odot}$ in TNG50, $7.5 \times 10^{6}\, \mathrm{M}_{\odot}$ in TNG100, and $5.9 \times 10^{7}\, \mathrm{M}_{\odot}$ in TNG300, with corresponding baryonic particle masses $8.5 \times 10^{4}\, \mathrm{M}_{\odot}$, $1.4 \times 10^{6}\, \mathrm{M}_{\odot}$, and $1.1 \times 10^{7}\, \mathrm{M}_{\odot}$, respectively. The spatial resolution improves from around 1 kpc in TGN300 to around $100-200$ pc in TNG50. This study primarily focuses on TNG100 for several reasons. First, TNG100 has been calibrated to reproduce key properties of observed galaxy populations. Second, the volume should be relatively large to contain a substantial number of groups, clusters and filaments, while the resolution is adequate to resolve low-mass galaxies. To assess the potential effects of resolution and cosmic variance, we compare our main results from TNG100 with those from TNG50 and TNG300 in Section 4.

\subsection{Galaxy samples and properties}\label{sec:property definition}
Galaxies with stellar masses of $M_{\ast} > 10^{8}\,\mathrm{M_\odot}$ in TNG100, $M_{\ast} > 10^{9}\, \mathrm{M_\odot}$ in TNG300, and $M_{\ast} > 10^{7}\, \mathrm{M_\odot}$ in TNG50 are selected from their respective catalogs, ensuring that each galaxy contains at least 100 star particles. At $z=0$, the TNG100 catalog includes a total of 49,914 galaxies with $M_{\ast} > 10^{8}\, \mathrm{M_\odot}$, comprising 27,729 central galaxies and 22,185 satellites. The key galaxy properties analyzed in this study are detailed below.

\begin{itemize}
\item  sSFR: the specific SFR, calculated by ${\mathrm{SFR}}/M_{\ast}$, using data from subhalo catalogs. 
\end{itemize}

\begin{itemize}
\item  Gas fraction: the ratio of gas mass bounded to subhalo, $M_\mathrm{gas}$, to the total baryonic mass in the galaxy, defined as $M_\mathrm{gas}/(M_\mathrm{gas}+M_{\ast})$. Data from subhalo and particles catalogs are used.
\end{itemize}

\begin{itemize}
\item  Cold Gas Fraction: indicates the abundance of raw materials for star formation and is defined as $M_\mathrm{cold}/(M_\mathrm{gas}+M_{\ast})$,  where $M_\mathrm{cold}$ is the mass of gas $T_\mathrm{gas} <\, 8000 \mathrm{K}$, that is lower than the typical temperature of warm neutral hydrogen (e.g. \citealt{2003ApJ...587..278W}), and $T_\mathrm{gas}$ is calculated by:
\begin{equation}
    T_\mathrm{gas} = (\gamma-1)*U/k_{\mathrm{B}}*\mathrm{m}_\mathrm{H}*\mu
\end{equation}
where $\gamma=5/3$, $\mu = 1$, $U$ represents internal energy, $k_\mathrm{B}$ is the Boltzmann constant, and $\mathrm{m_{H}}$ is the mass of hydrogen atom. Data from subhalo and particles catalog are used.
\end{itemize}

\begin{itemize}
\item $g - i$: galaxy color index derived from the absolute magnitudes of the u and i bands in AB system. For more information, see \citep{nelson2019illustristng}.
\end{itemize}

\begin{itemize}
\item  Mass-Weighted Age: calculated from stellar particles in subhalos.
\end{itemize}

\begin{itemize}
\item  Mass-Weighted stellar metallicity: calculated as,

\begin{equation}
    [\mathrm{M} / \mathrm{H}]_{\mathrm{M}}=\frac{\sum M_{\ast}[\mathrm{M} / \mathrm{H}]_{\ast}}{\sum M_{\ast}}
\end{equation}
where $M_{\ast}$ and $[\mathrm{M} / \mathrm{H}]_{\ast}$ represents the mass and metallicity of any stellar particle in the galaxy and $[\mathrm{M} / \mathrm{H}] = \mathrm{log}_{10}(Z/Z_{\odot} )$, with $Z_{\odot} = 0.0127$ in IllustrisTNG.
\end{itemize}

\begin{itemize}
\item log(O/H)+12: SFR-weighted gas phase metallicity calculated from star-forming gas particles in each galaxy. 
\end{itemize}

To control for the influences of halo mass and local environment, it is necessary to measure certain properties, including:

\begin{itemize}
\item $M_{\mathrm{h}}$: use the $M_{\mathrm{h}, \mathrm{C200}}$ from halo catalogs, which represents the total mass of the halo for galaxies enclosed within a sphere whose mean density is 200 times the critical density of the Universe.
\end{itemize}

\begin{itemize}
\item Overdensity: the local density contrast, $\delta_{\rho}$, is defined as: 
\begin{equation}
    \delta_{\rho} = \frac{\rho - \bar{\rho}}{\bar{\rho}}.
\end{equation}
We compute the dark matter particle density on a cubic grid covering the simulation volume using the CIC algorithm, with $35^3$ grids for TNG50, $75^3$ grids for TNG100, $205^3$ grids for TNG300. The density field is then smoothed using a Gaussian filter. In this process, we adopt a grid resolution and smoothing length of $1\, \mathit{c}\mathrm{Mpc/h}$, a value often used in observational studies (\citealt{ muldrew2012measures,o2024effect}). The galaxy overdensity is determined by interpolating the grid-based density at the galaxy's position. In the following context, we use $ 1 + \delta_{\rho}$ to indicate the local environment density. 
\end{itemize}

\subsection{Filaments identification and environment categories}
Cosmic filaments are identified using the DisPerSe algorithm, which detects the topological structure of the mass distribution (\citealt{sousbie2011persistent,sousbieet2011persistent}) and is widely used to extract filaments from galaxy and halo distributions. As input, we use the stellar mass and positions of galaxies with $M_{\ast} > 10^{9}\, \mathrm{M_\odot}$ {in each of the three TNG simulations, since many observations use galaxies with about $M_{\ast} > 10^{9}\, \mathrm{M_\odot}$ as tracers to identify filament (\citealt{tempel2014detecting,donnan2022role,o2024effect}).}

Due to the limited number of galaxies, we use the CIC algorithm instead of DTFE to generate a mass-weighted density field on the grids for processing with DisPerSe. The grid resolution and Gaussian smoothing scale are set to 0.547 $\mathit{c}\mathrm{Mpc/h}$ for TNG50, 0.585 $\mathit{c}\mathrm{Mpc/h}$ for TNG100 and 0.800 $\mathit{c}\mathrm{Mpc/h}$ for TNG300. The persistence threshold, -Nsig, is set to 4, and no additional smoothing is applied to the output skeleton. We have tested both the CIC and DTFE algorithms with various resolutions and -Nsig values. Based on visual inspection, we find that for our case (galaxy samples are used as input), filaments identified with density field produced by the CIC method would align more closely with the dark matter distribution compared to DTFE.

Given that a filament's influence on a nearby galaxy is primarily governed by the mass distribution in its immediate surroundings, we divide the filamentary structures identified by DisPerSe into relatively straight segments with lengths ranging from $1.5 \mathit{c}\mathrm{Mpc/h}$ and $2.5 \mathit{c}\mathrm{Mpc/h}$. Each of our segments can be roughly considered as a structure made up of several filament segments identified by DisPerSe, but with a cylindrical shape. The methods used to divide filaments into segments and measure their thickness, $R_{\mathrm{fil}}$, and mass, $M_{\mathrm{fil}}$, are briefly described in the Appendix. Further details on filament identification, segmentation, and thickness measurement will be presented in Yang et al. (in preparation). In this study, we include only filament segments with $M_{\mathrm{fil}} > 10^{12}\, \mathrm{M_\odot}$ and $R_{\mathrm{fil}} > 300\, \mathrm{ kpc} $ in our sample, as lower-mass, tenuous filaments are expected to have a negligible impact on galaxy properties.

\begin{figure}[htb]
\begin{center}
\hspace{-0.0cm}
\includegraphics[width=0.48\textwidth]{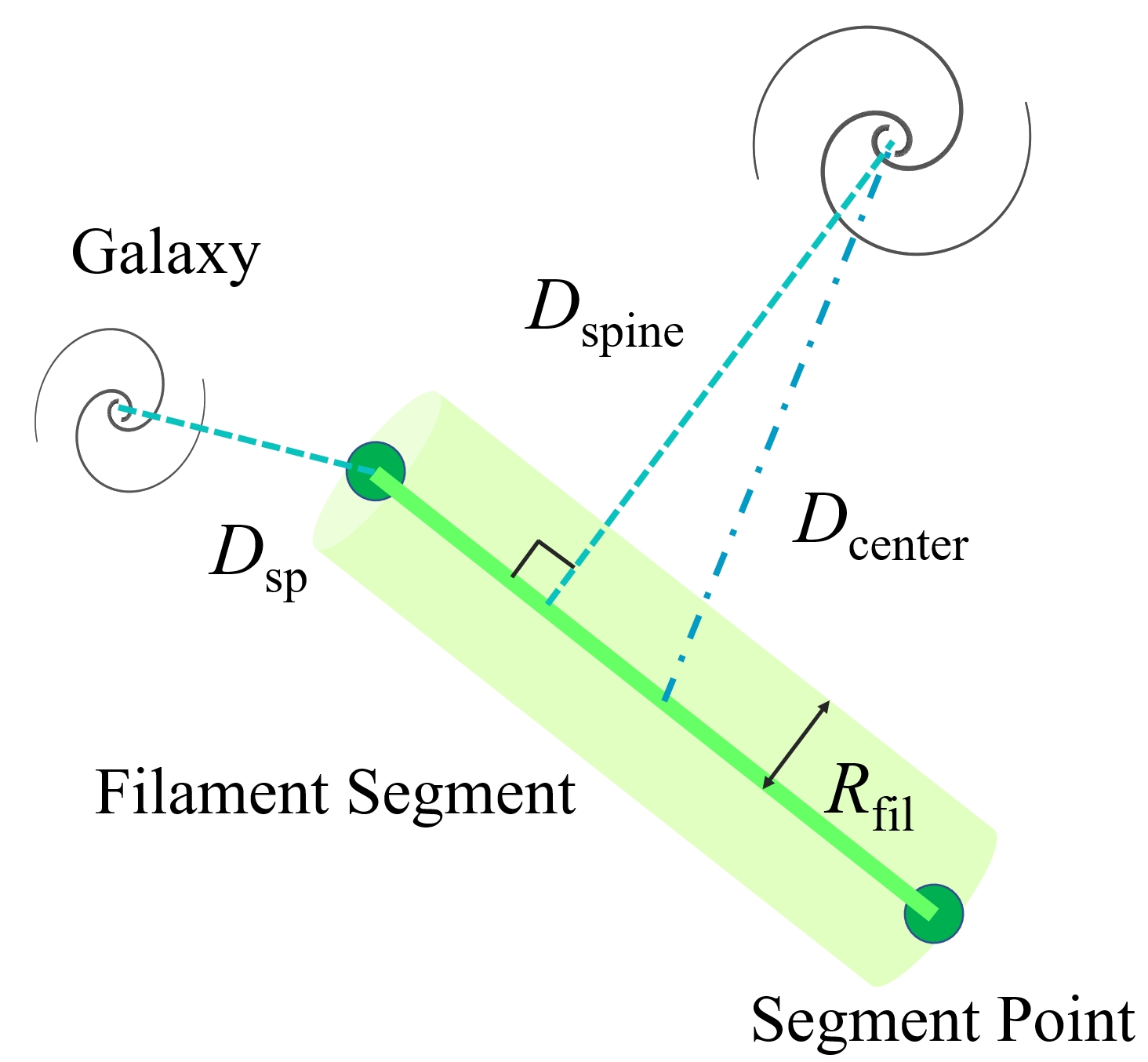}
\caption{A schematic representation of a filament segment and nearby galaxies. The filament segment is modeled as a cylindrical structure. The distance of a nearby galaxy to the filament spine, $D_{\mathrm{spine}}$, is defined as the perpendicular distance to the straight line connecting the two endpoints (depicted as a green-filled circle) of the segment  and is shown as a dashed line. The distance to the nearest segment endpoint, $D_{\mathrm{sp}}$, is also represented by a dashed line and corresponds to the distance from the galaxy to the closest endpoint. Additionally, the distance to the segment’s midpoint, $D_{\mathrm{center}}$, is illustrated as a dash-dot line.}
\label{fig:Dfil_defination}
\end{center}
\end{figure}

To examine the impact of filaments, we first measure the distance between a galaxy and its closest filament segment in our segment sample. We define this distance, $D_{\mathrm{fil}}$, according to their relative positions. As illustrated in Figure \ref{fig:Dfil_defination}, if the galaxy lies within an infinitely extended cylindrical region centered on the segment’s axis, $D_{\mathrm{fil}}$ is set to the perpendicular distance to the segment spine (axis) $D_{\mathrm{spine}}$. Otherwise, $D_{\mathrm{fil}}$ is defined as the distance to the nearest endpoint of the segment, $D_{\mathrm{sp}}$. This definition is chosen to avoid overestimating the influence of filaments when $D_{\mathrm{sp}}>D_{\mathrm{spine}}$. In TNG100, $53 \%$ of galaxies and their closest segments have $D_{\mathrm{fil}}=D_{\mathrm{spine}}$. For the remaining cases, $69 \%$ have $D_{\mathrm{sp}}<1.5 \times D_{\mathrm{spine}}$. We also tested an alternative definition of $D_{\mathrm{fil}}$ using the distance to the segment midpoint, denoted as $D_\mathrm{{center}}$ in Figure \ref{fig:Dfil_defination}. This alternative approach produces only minor changes in our results.

We classify the cosmic web environments of galaxies into four categories: Group, Field, Filament-Dominated Region (FilDR), and Group-Dominated Region (GrpDR). 

\begin{itemize}
\item The `Group' environment includes galaxies residing in halos with masses exceeding $M_{\mathrm{h}} > 10^{13}\, \mathrm{M_\odot}$. 

\item The `Field' environment is defined as the region sufficiently distant from groups and filaments, satisfying $D_{\mathrm{grp}} > 2 R_{\mathrm{grp}}$ and $D_{\mathrm{fil}} > 2 R_{\mathrm{fil}}$, where $D_{\mathrm{grp}}$ and $D_{\mathrm{fil}}$ denote the distances from a galaxy to its nearest group/cluster and nearest filament segment, respectively.

\item For galaxies in transitional regions between the Field and Group environments, we compare the gravitational potentials exerted by the nearest group/cluster and the nearest filament segment. Galaxies in regions where the filament’s potential dominates are classified as Filament-Dominated Regions (FilDR), while those where the group's potential is stronger are categorized as Group-Dominated Regions (GrpDR). 
\end{itemize}

These classification criteria are summarized in Table \ref{tab:environment_conditions}, along with the number of galaxies residing in each environment at $z=0$ in TNG100.

\begin{deluxetable*}{cccc}[htbp]
\tablenum{1}
\caption{Categories of cosmic web environments, their criteria, and number of central and satellite galaxies in these environments at redshift $z=0$ in the TNG100 simulation.}
\label{tab:environment_conditions}
\tablewidth{0pt}
\tablehead{
\colhead{Environment} & \colhead{Criteria} & \colhead{Number of centrals} & \colhead{Number of satellites}
}
\startdata
Group & In Groups and Clusters & 182 & 12138 \\
Field & $D_{\mathrm{grp}} > 2 R_{\mathrm{grp}}$ and $D_{\mathrm{fil}} > 2 R_{\mathrm{fil}}$ & 5489 & 772 \\
GrpDR & Not in Group or Field, and $|\mathit{\Phi}_{\mathrm{group}}| > |\mathit{\Phi}_{\mathrm{fil}}|$ & 11267 & 2778 \\
FilDR & Not in Group or Field, and $|\mathit{\Phi}_{\mathrm{fil}}| > |\mathit{\Phi}_{\mathrm{grp}}|$ & 10791 & 6497 \\
\enddata
\end{deluxetable*}

\begin{figure}[htb]
\begin{center}
\hspace{-0.0cm}
\includegraphics[width=0.48\textwidth]{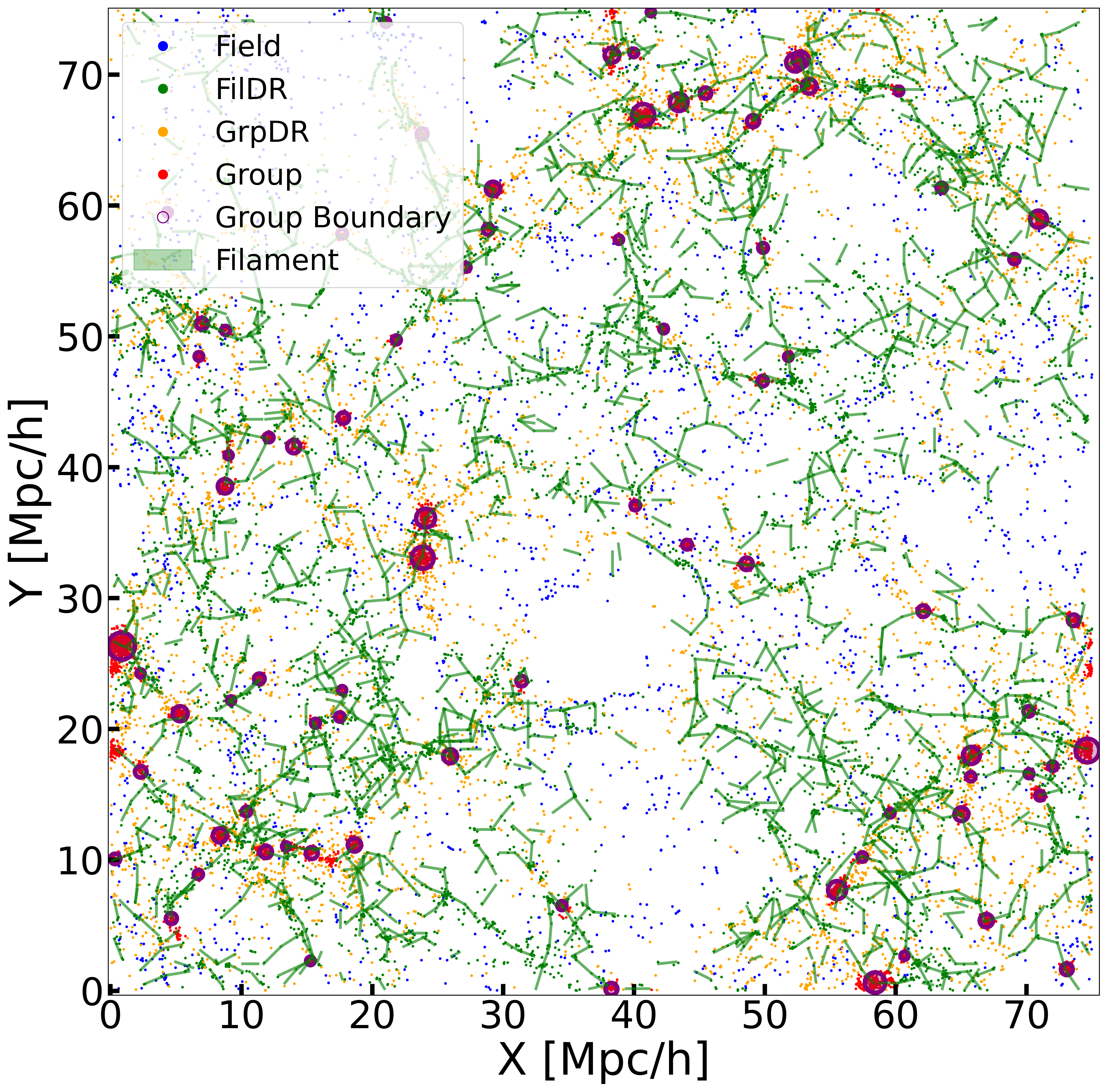}
\caption{Projected distribution of galaxy (dots) and filament samples within a slice of thickness $30\, \mathrm{Mpc/h}$ in TNG100. The environments in which galaxies residing are indicated by the colors: Field (blue), FilDR (filament-dominated regions, green), GrpDR (group-dominated regions, orange), and Group (red). Filaments are illustrated as green lines. Groups and clusters are denoted by purple circles, with radii corresponding to $R_{\mathrm{h}, \mathrm{Crit200}} $.}
\label{fig:galaxy_distribution_env}
\end{center}
\end{figure}

\begin{figure}[htb]
\begin{center}
\hspace{-0.0cm}
\includegraphics[width=0.45\textwidth]{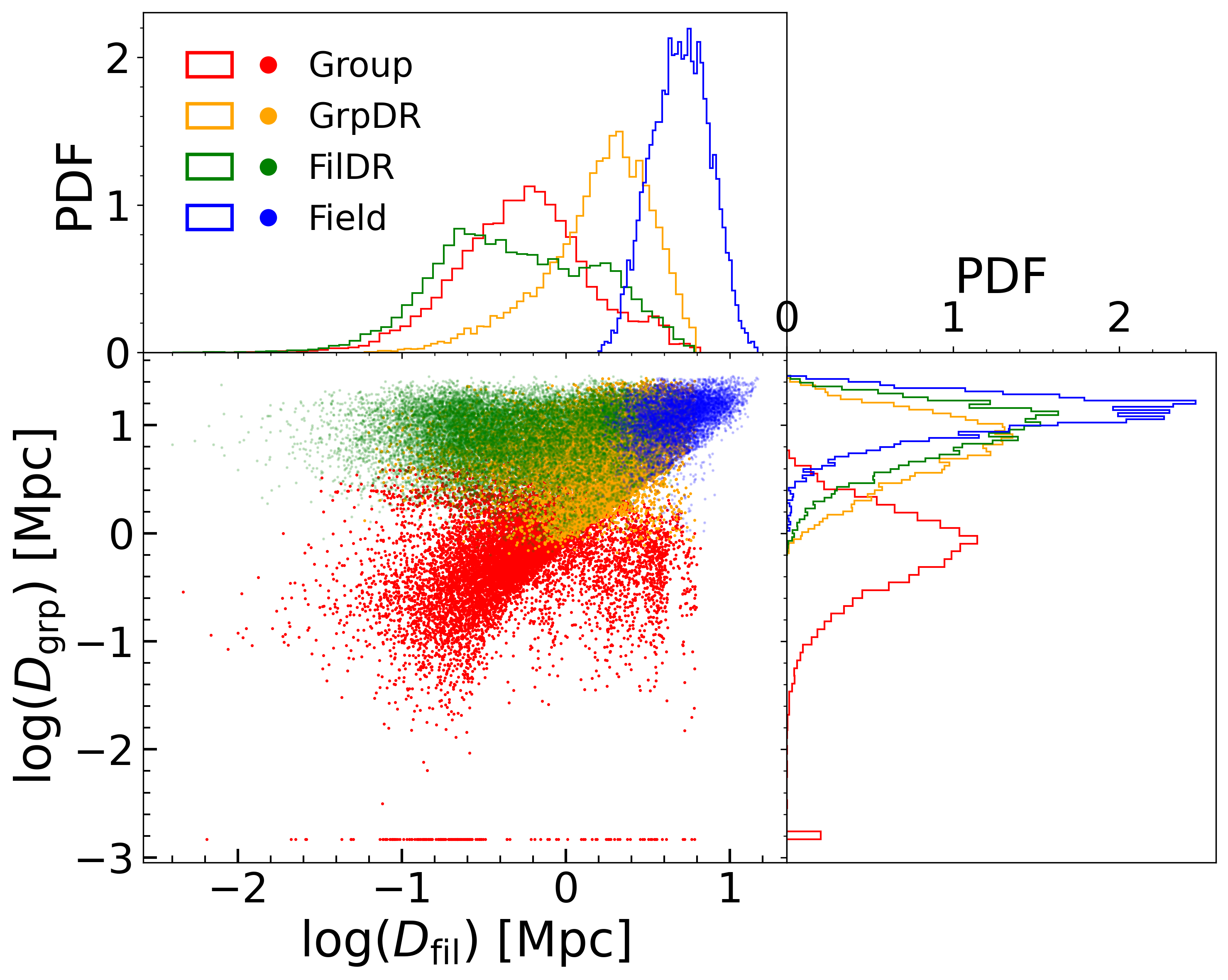}
\caption{The main plot shows the distribution of galaxies (dots) at $z=0$ in the TNG100 simulation in the distance to group versus distance to filaments (\( D_{\mathrm{grp}} \)–\( D_{\mathrm{fil}} \)) space. Red, orange, green and blue colors indicates environments of Group, GrpDR, FilDR, and Field, respectively. The top and right plots display the probability density function of \( D_{\mathrm{fil}} \) and \( D_{\mathrm{grp}} \), respectively, for galaxies in each environment.}
\label{fig:galaxy_distribution_D}
\end{center}
\end{figure}

\begin{figure}[htb]
\begin{center}
\hspace{-0.0cm}
\includegraphics[width=0.48\textwidth]{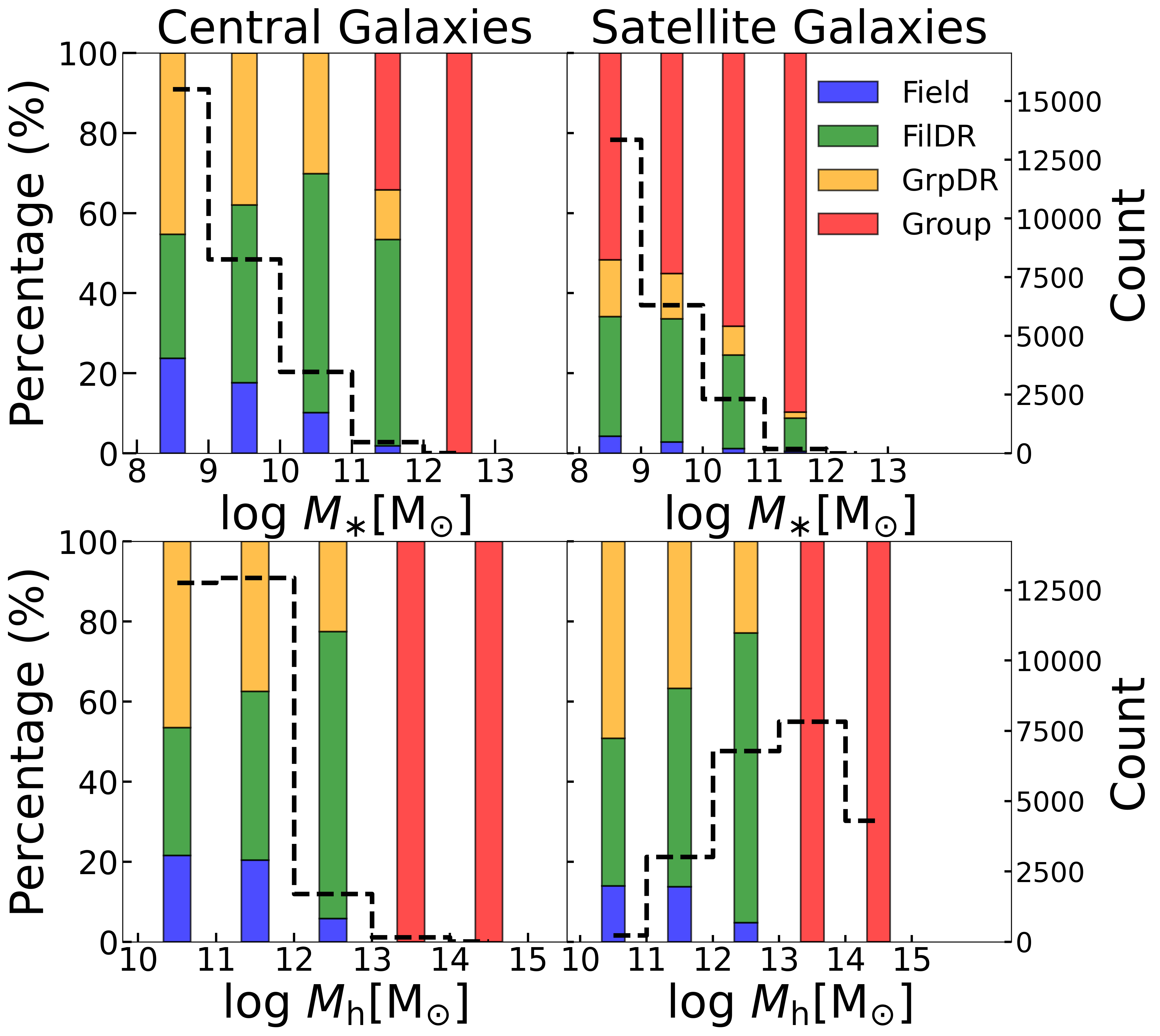}
\caption{The fraction of cental galaxies (left) and satellite galaxies (right) in four types of environments in different galaxy stellar mass bins (top) and halo mass bins (bottom), each with a bin width of 1 dex. Dashed lines represent the number of central or satellite galaxies corresponding to each mass bin.}
\label{fig:central_satellite_combined}
\end{center}
\end{figure}

Figure \ref{fig:galaxy_distribution_env} illustrates the projected distribution of galaxies (dots) within a $30\, \mathrm{Mpc/h}$ thick slice of TNG100. The colors indicate the different environments in which the galaxies reside. Filaments, depicted by green lines, act as bridges connecting massive galaxy groups and clusters while also linking relatively less massive galaxies. Notably, some galaxies appear to be in group-dominated regions despite being visually distant from any groups or clusters. This is because their nearest groups and clusters lie outside the selected slice.

Figure \ref{fig:galaxy_distribution_D} presents the distribution of galaxies in the parameter space defined by their distances from the nearest group, $D_{\mathrm{grp}}$, and the nearest filament segment, $D_{\mathrm{fil}}$. Galaxies in FilDR tend to be closer to filaments and farther from groups compared to those in GrpDR. This trend arises because filament segments generally contain less mass than groups and clusters, requiring galaxies to be in closer proximity for filaments to exert a significant gravitational influence. Additionally, the distribution of FilDR galaxies in $D_{\mathrm{fil}}$ is more extended with a slight bimodal feature. Notably, there is substantial overlap between FilDR and GrpDR in both $D_{\mathrm{fil}}$ and $D_{\mathrm{grp}}$, indicating that proximity alone does not uniquely determine whether filaments or groups have a dominant effect. The mass of the nearest filament and group must also be considered.

Figure \ref{fig:central_satellite_combined} presents the distributions of central and satellite galaxies across different stellar and halo mass bins in various environments for the $z=0$ sample in TNG100. Significant differences emerge between central and satellite galaxies. Approximately  $25\%$, $30\%$ and $45\%$ of the central galaxies with stellar mass below $10^{9.0}\,\rm{M_{\odot}}$ are located in the Field, FilDR, and GrpDR environments, respectively. As stellar mass increases from $10^{8.5}\,\rm{M_{\odot}}$ to $10^{11.5}\,\rm{M_{\odot}}$, the fraction of central galaxies residing in FilDR gradually rises, while the fractions in the Field and GrpDR decrease. All central galaxies with masses exceeding $10^{12.0}\,\rm{M_{\odot}}$ are found exclusively in groups and clusters. Additionally, the majority of central galaxies are hosted by halos with masses below $10^{13.0}\, \mathrm{M_\odot}$.

In contrast, the majority of satellite galaxies are found in groups and clusters, preferentially residing in more massive halos compared to their central counterparts. Additionally, the distribution of central galaxies provides insight into the spatial distribution of dark matter halos across different environments. Around $25\%$, $30\%$ and $45\%$ of halos with masses below $10^{11.0}\,\rm{M_{\odot}}$ are distributed among the Field, FilDR, and GrpDR regions, respectively. For halos with $M_\mathrm{h} < 10^{13.0}\,\rm{M_{\odot}}$, the fraction residing in FilDR gradually increases with mass, while those in the Field and GrpDR decrease. Since both stellar mass ($M_\ast$) and halo mass ($M_\mathrm{h}$) play crucial roles in shaping galaxy properties, this variation complicates efforts to simultaneously control for both factors across the entire sample. To address this, we will first examine the impact of the cosmic web and filaments on central and satellite galaxies separately in the following section.

\section{Impact of cosmic web on galaxies properties in TNG100}\label{sec:results}

\subsection{Central galaxies}\label{sec:central_galaxies}

\begin{figure*} 
\begin{centering}
\includegraphics[width=0.95\textwidth]{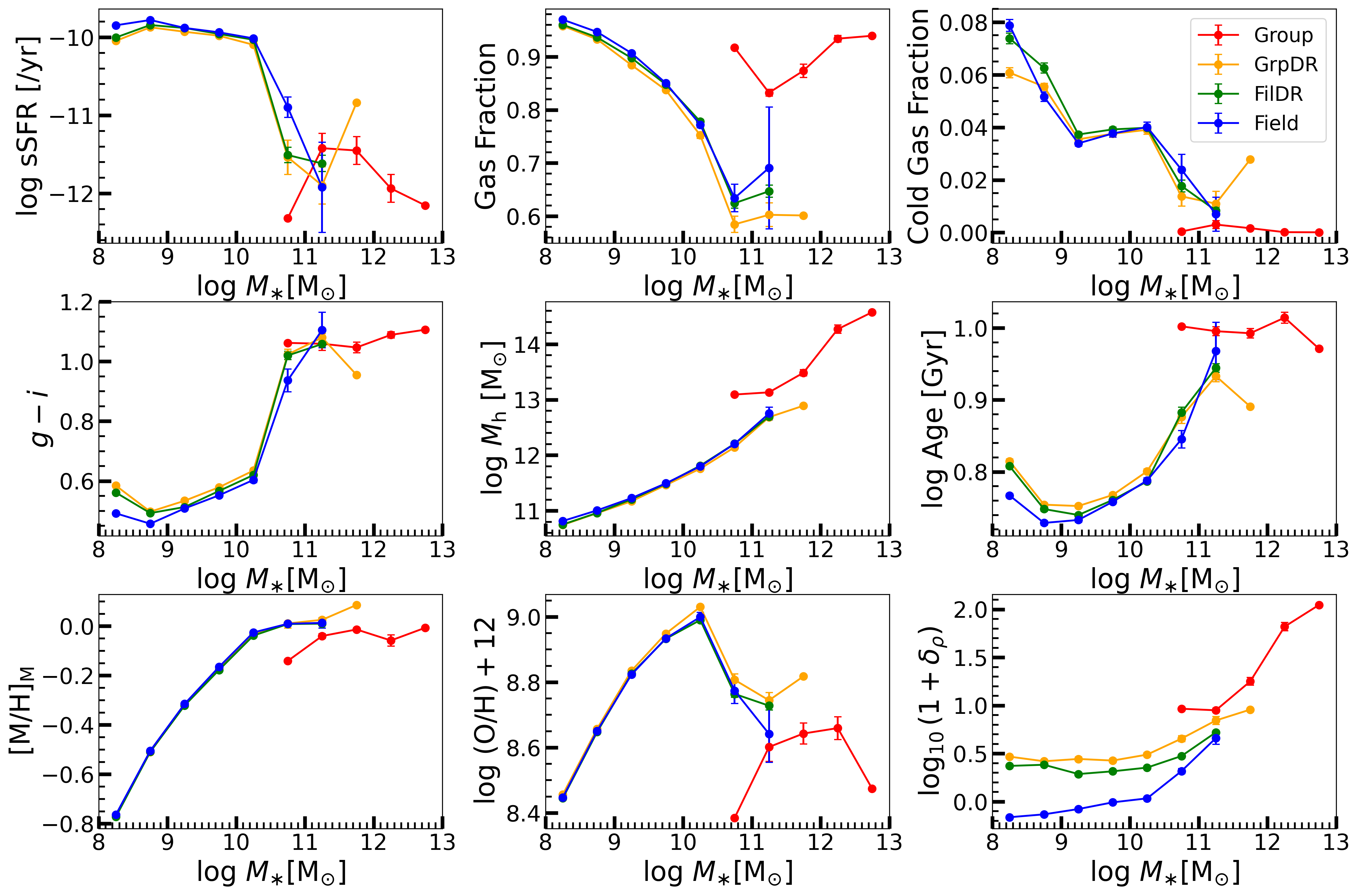}
\caption{Galaxy properties as functions of the stellar masses for central galaxies in Group (red), GrpDR (orange), FilDR (green), Field (blue). The panels from left to right and from top to bottom show the median sSFR, gas fraction, cold gas fraction, color index $(g - i)$, halo mass, mass-weighted stellar age, mass-weighted stellar metallicity ($[\mathrm{M} / \mathrm{H}]_{\mathrm{M}}$), and overdensity. The definitions of the above properties are given in section \ref{sec:property definition}. Error bars represent $1 \sigma$, calculated by standard deviation.}
\label{fig:porperties_central}
\end{centering}
\end{figure*}

Figure \ref{fig:porperties_central} shows the properties of central galaxies as functions of stellar masses across different environments. The median value and $1 \sigma$  uncertainties for each stellar mass bin are estimated using 200 bootstrap resampling iterations (\citealt{efron1992bootstrap,bootstrap}). Overall, most galaxy properties exhibit a clear dependence on stellar mass, consistent with previous studies and observations (\citealt{nelson2018first,2018MNRAS.473.4077P,pillepich2018first,springel2018first,torrey2019evolution}). For central galaxies with $\rm{M_{\ast}} < 10^{10.25}\, \mathrm{M_\odot}$, the majority are blue and star-forming, maintaining a reservoir of cold gas that sustains star formation. The metallicity of both the gas and stellar components increases with $\rm{M_{\ast}}$, driven by enrichment from stellar winds and supernovae. However, for galaxies exceeding $\rm{M_{\ast}} > 10^{10.25}\, \mathrm{M_\odot}$, their specific star formation rate (sSFR) declines rapidly, they become progressively redder, and their stellar ages increase significantly. These trends stem from two key factors. First, cold gas accretion onto the halo becomes inefficient when the halo mass exceeds $10^{12}\, \mathrm{M_\odot}$. Second, AGN feedback becomes increasingly effective in suppressing gas cooling in both the circumgalactic medium (CGM) and the interstellar medium (ISM), ultimately leading to the quenching of star formation in galaxies with $\rm{M_{\ast}} > 10^{10.25}\, \mathrm{M_\odot}$ (\citealt{zinger2020ejective,piotrowska2022quenching}).

For central galaxies of a given stellar mass, their host halos have almost the same mass, regardless of the various environments. Beyond halo mass, the cold gas fraction, stellar metallicity, and log(O/H)+12 exhibit only minor variations between galaxies in the Field, FilDR, and GrpDR, aside from fluctuations within specific mass ranges. However, noticeable differences emerge in sSFR, gas fraction, color, age, and overdensity for galaxies with $M_{\ast} < 10^{10.25}\, \mathrm{M_\odot}$ across different environments. These results indicate that while stellar mass and halo mass primarily govern the properties of central galaxies, as demonstrated by \cite{alpaslan2015galaxy}, superhalo-scale environmental effects also play a role. Furthermore, the impact of these superhalo-scale environments is more pronounced for low-mass galaxies, suggesting that they are particularly susceptible to external environmental influences.

Specifically, within the stellar mass range of $10^{8.25}\, \mathrm{M_\odot}$ to $10^{10.25}\,\mathrm{M_\odot}$, central galaxies in the Field environment exhibit higher sSFR and gas fractions, bluer colors, younger stellar populations, and lower local densities compared to their counterparts in FilDR. A similar trend is observed when comparing galaxies in FilDR to those in GrpDR, but the differences are smaller. These findings align with general observational trends (e.g., \citealt{2004ApJ...617...50R,2007ApJ...658..898P,2012MNRAS.426.3041H,2021ApJ...906...97F,dominguez2023galaxies}). It is worth noting that many previous studies do not distinguish between central and satellite galaxies. In Section \ref{sec:compare_observation}, we extend our analysis to include both populations for a more comprehensive comparison with observations.

Quantitatively, while the differences in sSFR, gas fraction, color, and stellar age are relatively moderate, whereas the variation in overdensity is more pronounced. For example, at a stellar mass of $M_{\ast} = 10^{8.25}\, \mathrm{M_\odot}$, galaxies in the Field environment exhibit an sSFR higher than those in FilDR by $0.16 \pm 0.015$ ($11.0\sigma$). This difference remains significant at $M_{\ast} = 10^{8.75}\,\mathrm{M_\odot}$, with an sSFR offset of $0.065 \pm 0.013$ ($5.0\sigma$), but disappears by $M_{\ast} = 10^{10.25}\, \mathrm{M_\odot}$, where the difference is only $0.017 \pm 0.023$ ($0.74\sigma$). Meanwhile, the disparity in sSFR between GrpDR and the other environments remains distinguishable
 at intermediate stellar masses. At $M_{\ast} = 10^{8.25}\, \mathrm{M_\odot}$, galaxies in GrpDR have an sSFR lower than those in FilDR by $0.044 \pm 0.017$ ($2.6\sigma$), and at $M_{\ast} = 10^{10.25}\, \mathrm{M_\odot}$, the difference is $0.065 \pm 0.025$ ($2.6\sigma$). On average, for central galaxies with $M_{\ast} < 10^{10.5}\,\mathrm{M_\odot}$, the sSFR difference is $0.053$ between the Field and FilDR, and $0.042$ between FilDR and GrpDR.

It is important to note that we distinguish the relative influence of groups and filaments based on gravitational potential. Some galaxies classified as being in FilDR are actually located in proximity to groups and clusters. Previous studies have shown that groups and clusters can exert significant effects on nearby galaxies (e.g., \citealt{winkel2021imprint, donnan2022role}). 
In TNG100, approximately $1\%$ of galaxies in FilDR and $6\%$ of galaxies in GrpDR lie within $2 \times R_{200c}$ of their nearest group or cluster.
The observed differences in galaxy properties between the Field and FilDR, as well as between FilDR and GrpDR, as shown in Figure \ref{fig:porperties_central}, may be partially influenced by these galaxies. To assess this potential effect, we reanalyzed the results after excluding galaxies within $2 \times R_{200c}$ of groups and clusters. While the main findings remain unchanged, the differences between galaxies in FilDR and GrpDR narrow slightly, as shown in Figure \ref{fig:property_StellarMass_central_TNG100_out2R}.

\begin{figure}[htb]
\begin{center}
\hspace{-0.0cm}
\includegraphics[width=0.45\textwidth]{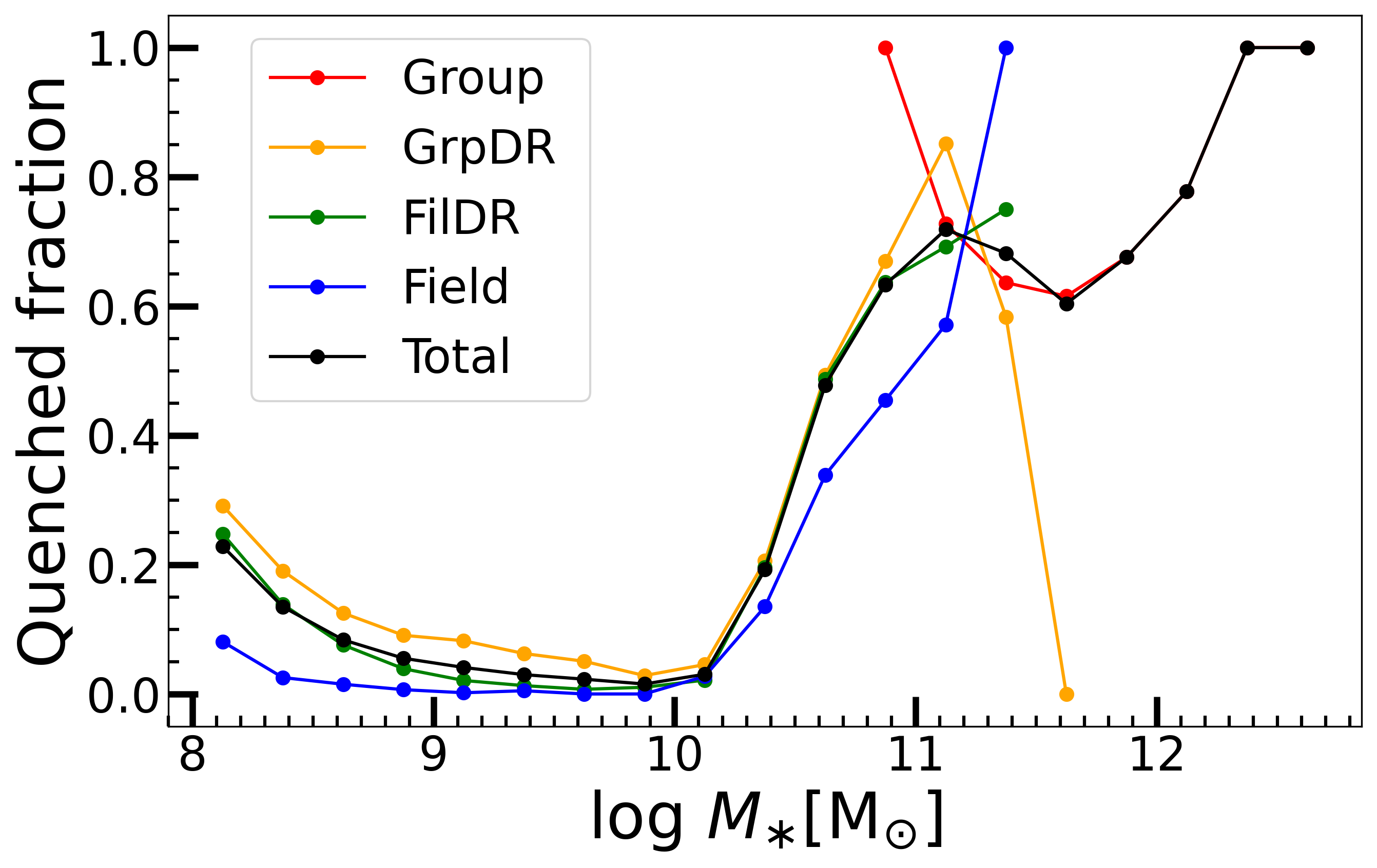}
\caption{The quenched fractions of central galaxies as a function of stellar mass. The star forming or quenched state is inferred from the $(g - z)$ color and r band absolute magnitude. Different colors indicate quenched fractions in various environments: Group (red), GrpDR (orange), FilDR (green), Field (blue), and total (black).}
\label{fig:quenched_fraction_central}
\end{center}
\end{figure}

Furthermore, we investigate the quenched fraction across different environments, a topic frequently discussed in the literature. Rather than using a fixed sSFR threshold, we adopt the $(g - z)$ color index, which has been shown to provide a clearer separation between quenched and star-forming galaxies compared to the widely used $(g - i)$ color, particularly for faint galaxies (\citealt{meng2023galaxy}). 
The color-based classification of all galaxies is provided in the Appendix. The demarcation line separating star-forming and quenched galaxies is as follows:
\begin{equation}
    g - z = 0.95 - 0.3 \tanh\left(\frac{M_{\mathrm{r}} + 15}{5}\right)
\end{equation}
This form is derived from \citealt{meng2023galaxy}, while the parameters have been determined according to KDE density map of g-z-Mr. Figure \ref{fig:quenched_fraction_central} illustrates the quenched fraction of central galaxies as a function of stellar mass in various environments.

Our results show that GrpDR hosts a higher fraction of quenched galaxies with $\rm{M_{\ast}}<10^{10.75}\, \mathrm{M_\odot}$ compared to FilDR and Field, with the differences becoming more pronounced at lower masses. This trend aligns with the sSFR distribution and suggests that galaxies in transition regions may experience a sequence of environmental effects, first undergoing pre-processing by filaments before being influenced by groups, before eventually falling into dense group environments, as interpreted by \cite{kuchner2022inventory}. Meanwhile, the quenched fraction of central galaxies increases as the stellar mass decreases at the low mass end ($\rm{M_{\ast}} < 10^9\,\mathrm{M_\odot}$). This is mainly because the TNG simulations produce an excess of quenched low-mass galaxies with $\rm{M_{\ast}} < 10^9\,\mathrm{M_\odot}$ compared to SDSS 7 (\citealt{2021MNRAS.502.1051A}). We have also tested to use a threshold of $\mathrm{log}(sSFR) = -11$ to separate star forming and quenched galaxies, and the result is consistent with current one shown in Figure \ref{fig:quenched_fraction_central}.

On the other hand, central galaxies in the Group environment exhibit lower sSFR and cold gas fractions, higher stellar metallicities, redder colors, older stellar populations, and greater local overdensities compared to those in non-Group environments. These differences are primarily driven by the massive halos hosting these galaxies. However, since central galaxies in the Group environment tend to have significantly higher stellar and halo masses, it is not possible to directly compare them to galaxies in other environments while controlling for these two factors.

\begin{figure}[htb]
\begin{center}
\hspace{-0.0cm}
\includegraphics[width=0.45\textwidth]{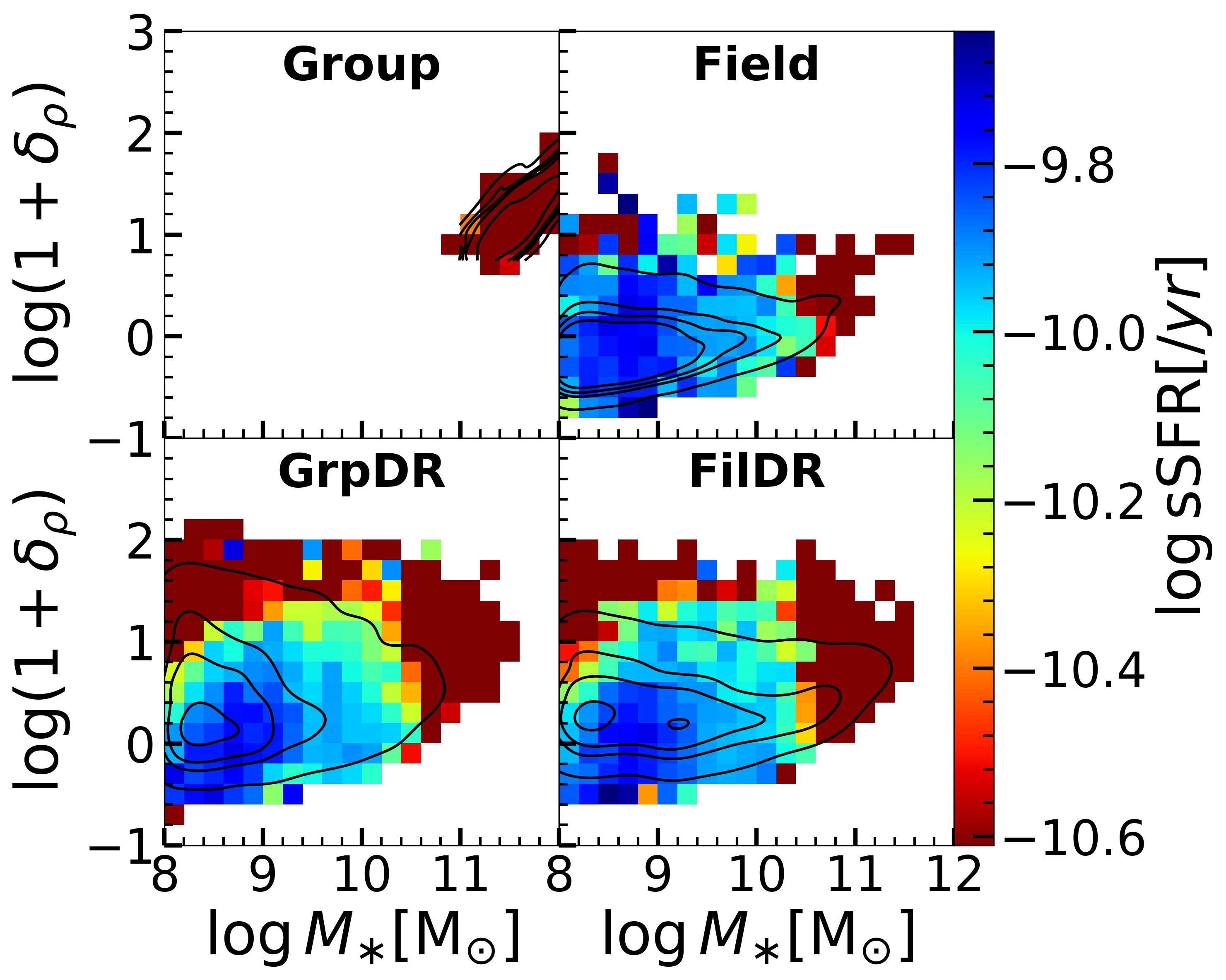}
\caption{Heatmap of sSFR for central galaxies: Overdensity vs. Stellar Mass. Each pixel represents the median sSFR of galaxies within a specific stellar mass and local overdensity bin of 0.2 dex. The sSFR values are indicated by the color bar on the right. Contours, from the innermost to the outermost, enclose 5\%, 25\%, 45\%, 65\%, and 85\% of the probability density function (PDF) in the overdensity-stellar mass parameter space, calculated using Kernel Density Estimation (KDE).
}
\label{fig:overdensity_correction_central}
\end{center}
\end{figure}

\begin{figure*} 
\begin{centering}
\hspace{-0.0cm}
\includegraphics[width=0.95\textwidth]{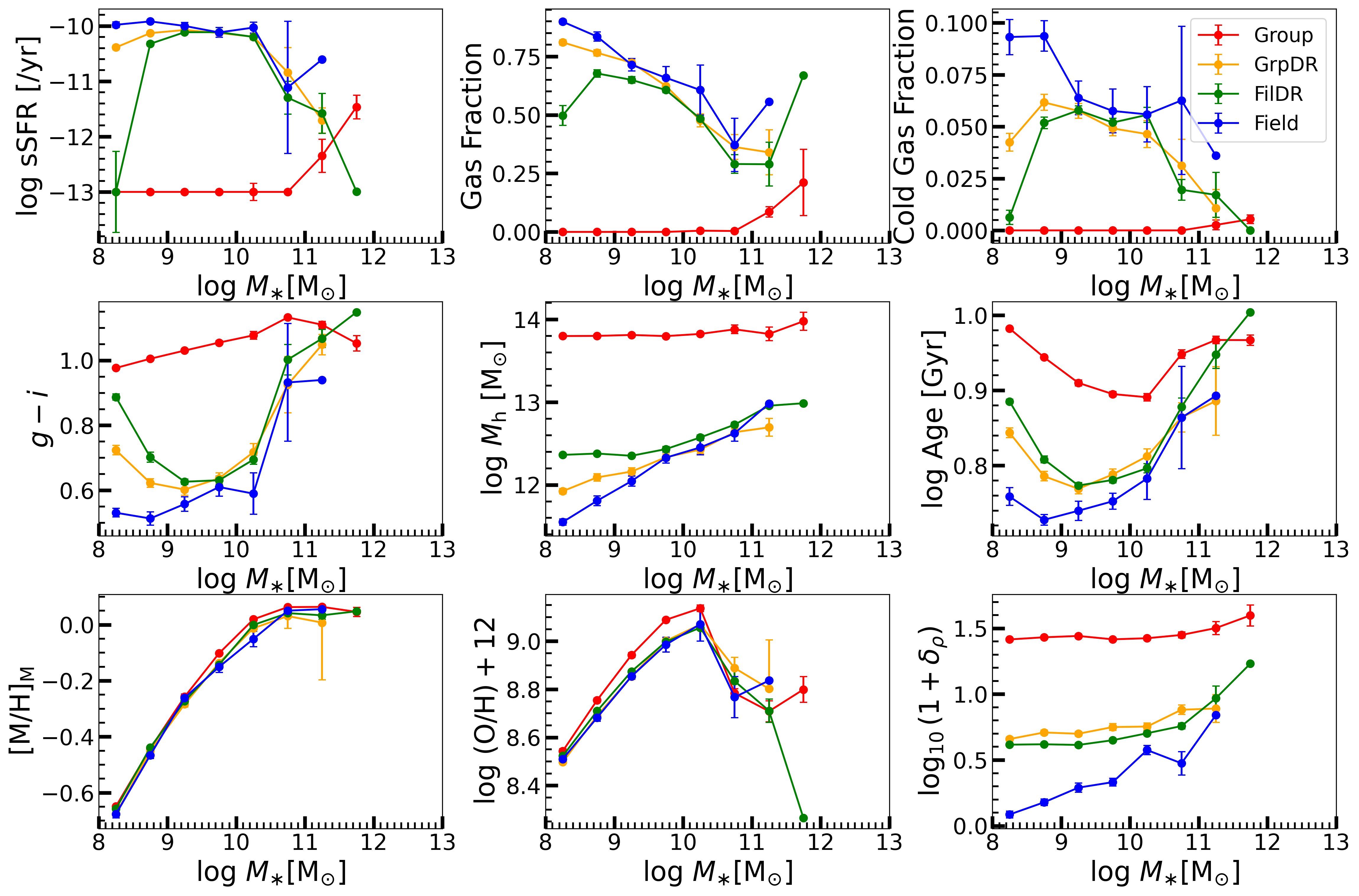}
\caption{Same as Figure \ref{fig:porperties_central}, but for satellite galaxies.}
\label{fig:porperty_satellite}
\end{centering}
\end{figure*}

\subsection{Overdensity correction for centrals}\label{sec:overdensity_correction}

Beyond the influence of stellar mass, halo mass, and group/cluster environments, it is also crucial to account for local density when isolating the impact of cosmic web filaments. To achieve this, we compare the sSFR of galaxies with similar stellar mass and overdensity across different environments. Figure \ref{fig:overdensity_correction_central} presents the results using a bin size of 0.2 dex. Next, we quantify the difference in median sSFR between environments for central galaxies within the same stellar mass and overdensity bins. To minimize statistical fluctuations due to limited sample sizes, we only include bins with at least five galaxies in both environments. When overdensity is controlled, the mean sSFR difference for central galaxies with $M_{\ast} < 10^{10.4}\, \mathrm{M_\odot}$ is -0.005 dex between the Field and FilDR environments and -0.005 dex between FilDR and GrpDR. In contrast, without controlling for overdensity, these differences are 0.053 dex and 0.042 dex, respectively. However, this comparison remains inapplicable to galaxies in groups and clusters, as their stellar and halo masses are significantly higher than those in other environments.

This suggests that local overdensity takes the vast majority of responsibility for the differences between central galaxies across various cosmic web environments. This finding partially aligns with the results of \cite{o2024effect}, and is consistent with previous studies on the correlation of galaxies properties and local overdensity (e.g. \citealt{2004MNRAS.353..713K,2018ARA&A..56..435W}). Moreover, it supports prior research showing that local environmental density becomes an effective quenching mechanism for galaxies at redshifts $<1$ (e.g. \citealt{2010ApJ...721..193P,popesso2011effect, sobral2011dependence,darvish2016effects}). Several physical mechanisms could contribute to the enhanced suppression of star formation in denser environments, including ram pressure stripping, starvation, galaxy-galaxy interactions, tidal interactions with clusters, an increased merger rate, preprocessing and cosmic web stripping (e.g. \citealt{herzog2023present,liao2019impact}). Additionally, differences driven by overdensity may be partially explained by the "archaeological downsizing" scenario (\citealt{thomas2005epochs}). Galaxies in high-density regions likely formed earlier than those in low-density regions due to accelerated proto-halo collapse in an enhanced density field. Consequently, high-density environments such as Groups, GrpDR, and FilDR should contain a higher fraction of older galaxies compared to the Field.

We also find that, when the effect of overdensity is controlled, the differences in color, gas fraction, and age between galaxies of similar stellar mass but in distinct environments follow a trend similar to that of sSFR. This suggests that cosmic web filaments exert a minor influence on the sSFR, color, gas fraction, and age of central galaxies in TNG100, beyond the effects of stellar mass, halo mass, groups and overdensity. 

\subsection{Satellite galaxies}\label{sec:satellite_galaxies}

\begin{figure}[htb]
\begin{center}
\hspace{-0.0cm}
\includegraphics[width=0.45\textwidth]{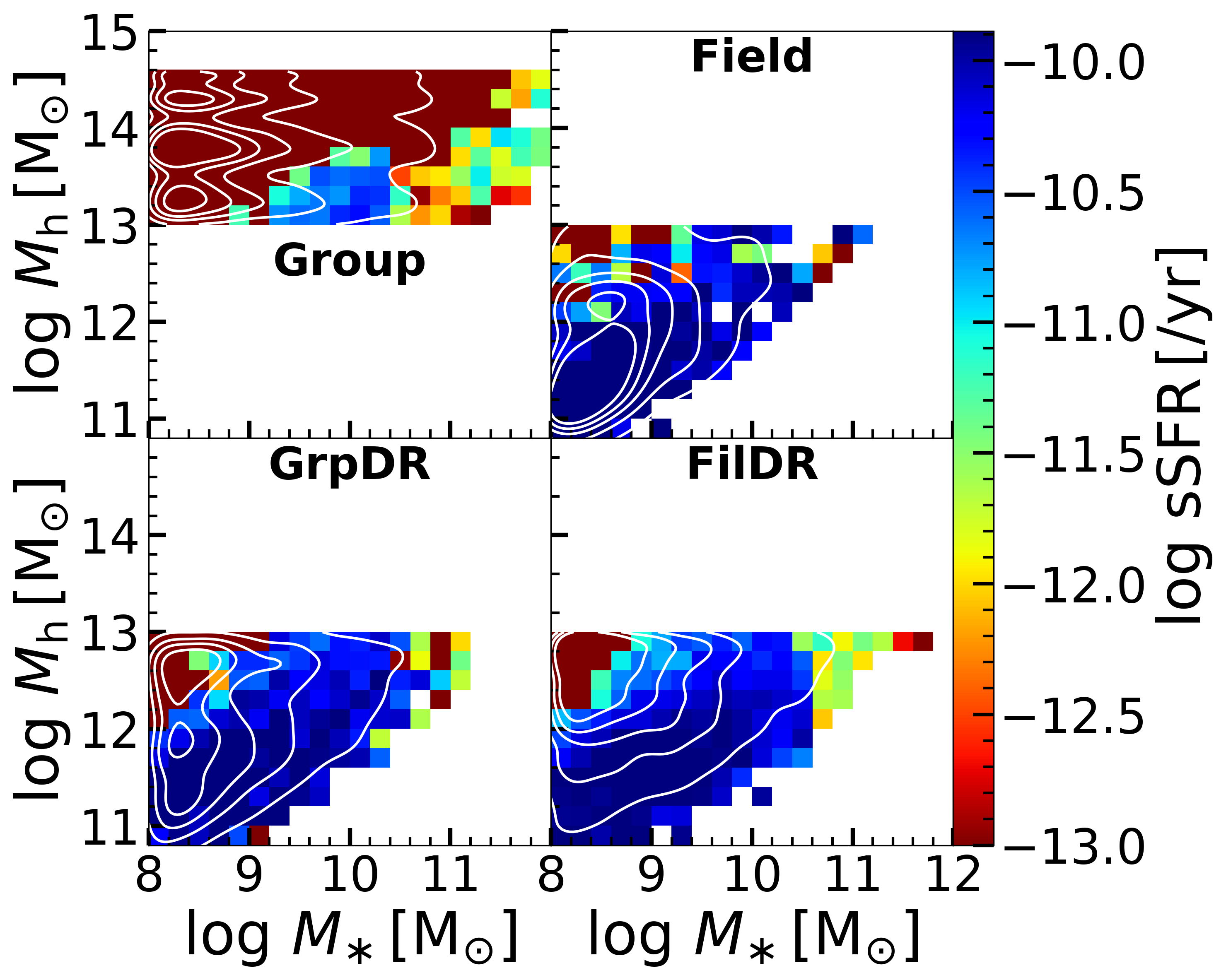}
\caption{Heatmap of sSFR for satellite galaxies: Halo mass vs. Stellar Mass. Each pixel represents the median sSFR of galaxies within a specific stellar mass and halo mass bin of 0.2 dex. The sSFR values are indicated by the color bar on the right. Contours, from the innermost to the outermost, enclose 5\%, 25\%, 45\%, 65\%, and 85\% of the PDF in the stellar mass-halo mass parameter space, calculated using Kernel Density Estimation (KDE).}

\label{fig:halomass_correction_satellite}
\end{center}
\end{figure}

\begin{figure}[htb]
\begin{center}
\hspace{-0.0cm}
\includegraphics[width=0.45\textwidth]{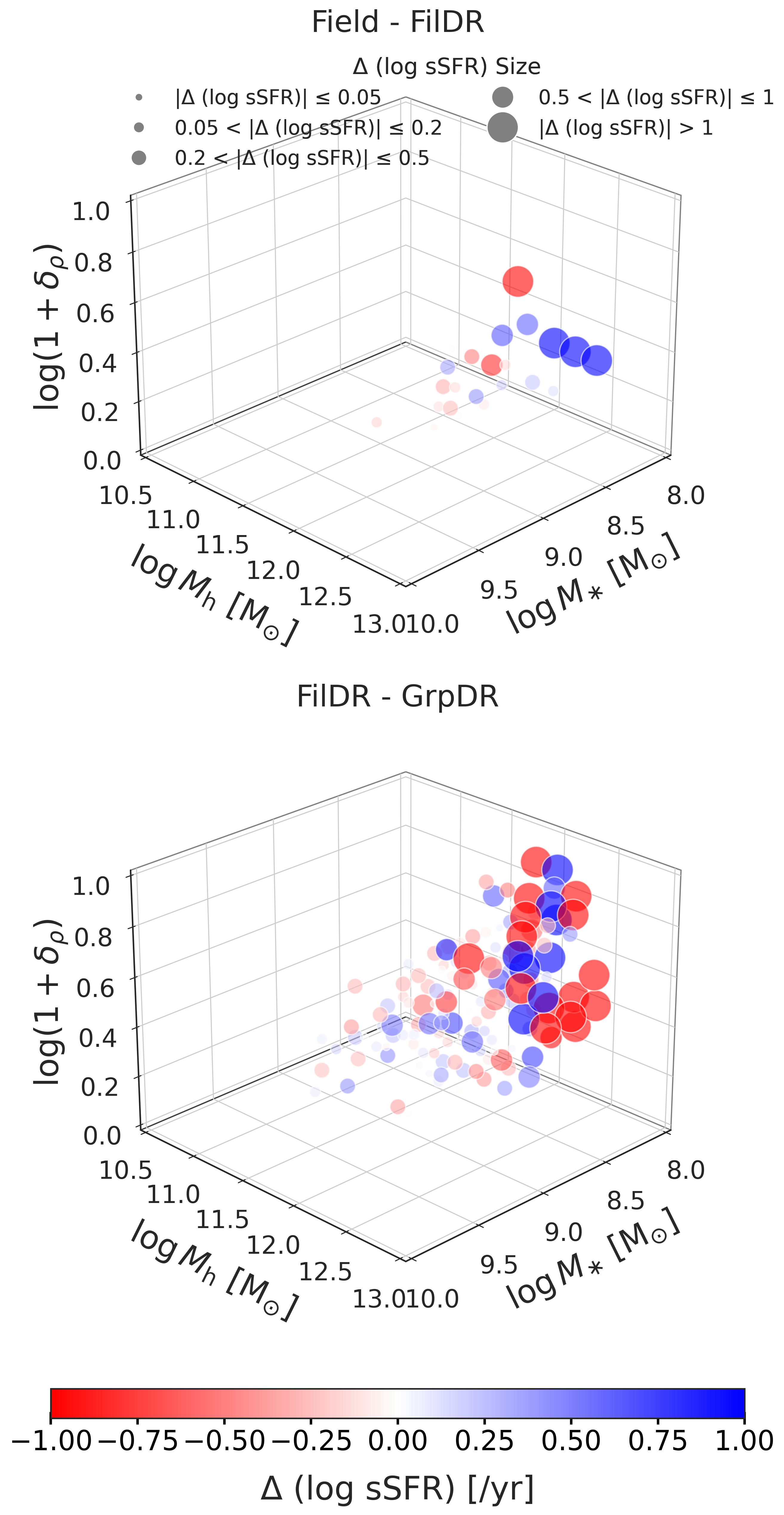}
\caption{Differences in sSFR between satellite in the same stellar mass - halo mass - overdensity bins but in different cosmic web environments. The upper (bottom) panel indicates the difference between Field and FilDR (FilDR and GrpDR). Each sphere represents the difference in the median sSFR within a specific bin that containing at least 5 samples in both environments. 
}
\label{fig:three_factor_correction_satellite}
\end{center}
\end{figure}

\begin{figure*} 
\begin{center}
\hspace{-0.0cm}
\includegraphics[width=0.95\textwidth]{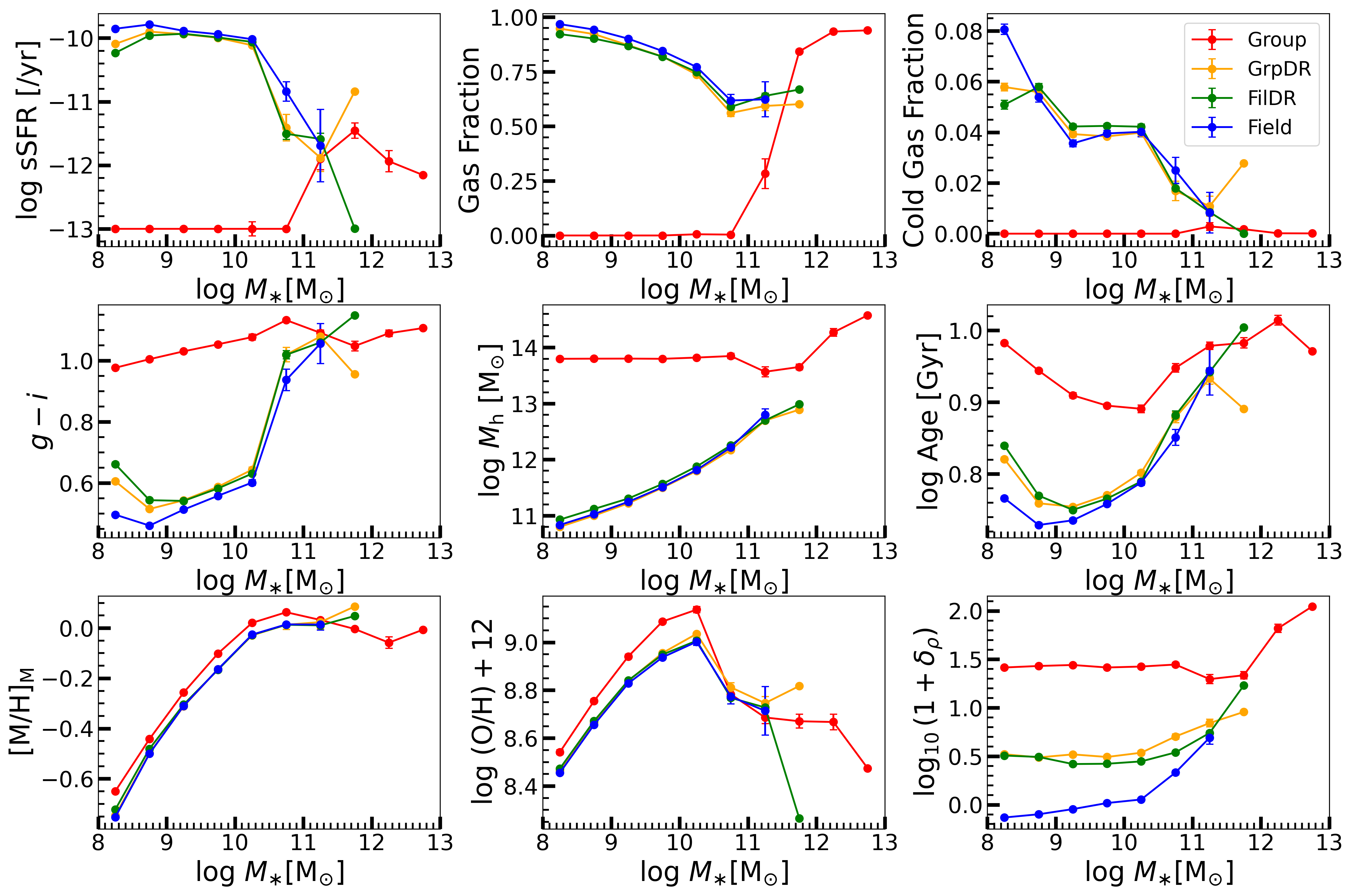}
\caption{Same as Figure \ref{fig:porperties_central}, but for the whole galaxy samples including both centrals and satellites in TNG100.}
\label{fig:porperty_together}
\end{center}
\end{figure*}

Figure \ref{fig:porperty_satellite} displays the properties of satellite galaxies as functions of stellar mass across different environments. Most properties, except for gas and stellar metallicity, show significant variations among satellites of the same stellar mass in different environments, which are more evident than the differences between centrals. The most pronounced differences arise between satellites in groups and those in the other three environments. Nearly all satellites in groups and clusters with stellar masses below $10^{10.75}\, \rm{M_{\odot}}$ are quenched, exhibiting very low sSFR, depleted gas reservoirs, higher host halo masses, and significantly older stellar populations. Compared to satellites in the Field, those in FilDR and GrpDR have lower sSFR, reduced gas and cold gas fractions, redder colors, more massive host halos, older stellar populations, and higher local densities. These features are broadly consistent with \cite{herzog2023present} and \cite{hasan2023evolving}, where satellites in dense cosmic environments exhibit significantly reduced star formation and lower gas fractions at low redshift. However, satellite galaxies in FilDR exhibit even lower sSFR and gas fractions, redder colors, more massive host halos, and older stellar populations than those in GrpDR, which is opposite to the relative trend observed for central galaxies. Nonetheless, similar to central galaxies, satellites in GrpDR experience higher local densities than those in FilDR.

The mass of the host halo plays a crucial role in shaping galaxy evolution, even when stellar mass is matched (e.g., \citealt{mo2010galaxy,2015MNRAS.447..374G,2018ARA&A..56..435W,wang2023environmental}). This effect is particularly pronounced for satellite galaxies (\citealt{2015ApJ...800...24K}) and is likely the primary driver of the differences observed among satellites with the same stellar mass across different environments that shown in Figure \ref{fig:porperty_satellite}. For instance, satellites with $M_{\ast} \leq 10.25 \, \rm{M_{\odot}}$ in FilDR reside in host halos with masses around $M_{\mathrm{h}} = 10^{12.4}\, \mathrm{M_\odot}$, which are systematically more massive than those in GrpDR and the Field. Additionally, the cold gas fraction of low-mass satellites ($\rm{M_{\ast}}\leq 10^{9.25}\,\rm{M_{\odot}}$) in FilDR and GrpDR increases with stellar mass, a trend that differs from that of central galaxies. This pattern may be attributed to the fact that low-mass satellites are more vulnerable to ram-pressure stripping due to their weaker gravitational restoring forces (\citealt{gunn1972infall,xie2020influence}).

To disentangle the effects of stellar mass and halo mass, we examine the difference in sSFR of satellite galaxies while controlling for both parameters. The results are presented in Figure \ref{fig:halomass_correction_satellite}. Note that, the colorbar in Figure \ref{fig:halomass_correction_satellite} has a much larger range ($\sim 3$ dex) than that of Figure \ref{fig:overdensity_correction_central} ($\sim 1$ dex), due to satellites exhibit a much larger range of star formation activity than centrals.
To minimize fluctuations due to limited sample size, we calculate the sSFR differences between environments only for bins containing at least five galaxies in both environments. When halo mass is accounted for, the mean sSFR difference between the Field and FilDR environments decreases significantly from 0.737 dex to 0.039 dex, while the difference between FilDR and GrpDR is reduced from -0.564 dex to 0.141 dex. These findings indicate that halo mass plays a dominant role in shaping the properties of satellite galaxies. However, the residual differences suggest that factors beyond stellar mass and host halo mass also influence satellite galaxy properties. The cosmic web environment may contribute to these variations, though additional factors, such as local density likely play a role as well.

As an example of further controlling the effect of local overdensity, Figure \ref{fig:three_factor_correction_satellite} illustrates the differences in sSFR for satellite galaxies with matched stellar mass, halo mass, and overdensity across three environments. When all three factors are taken into account, the residual difference between FilDR and GrpDR is -0.006 dex, However, when all three factors are matched, the mean sSFR difference between the Field and FilDR environments is 0.123 dex, which is larger than when only stellar mass and halo mass are matched. This increase may partly result from the reduced sample size, as we only compare bins containing at least five galaxies in both environments. Additionally, this change suggests that an additional factor is needed to fully explain the discrepancy. The influence of cosmic web filaments is a strong candidate, as suggested by previous studies (\citealt{kuutma2017voids, 2019MNRAS.483.3227K, song2021beyond}).  

The anisotropy of the cosmic web environment could impact galaxies through several potential physical mechanisms. For instance, the properties of host halos, such as formation time and mass accretion rate, can be influenced by the halo's position within the cosmic web (\citealt{2017JCAP...03..059L, 2018MNRAS.476.4877M}).
Additionally, processes like spin advection, that is galaxies acquire angular momentum from the large-scale vorticity of the anisotropic environment, inefficient gas transfer from halos to galaxies due to large angular momentum, and AGN feedback at high mass may all contribute to these differences (\citealt{2019MNRAS.483.3227K, song2021beyond}). Moreover, \cite{liao2019impact} indicates that cold and dense gas pre-processed by dark matter filaments can be further accreted into residing individual low-mass haloes in directions along the filaments. About 30 percent of the accreted gas of a residing filament halo was pre-processed by filaments, leading to two different thermal histories for the gas in filament haloes. Also, cosmic web stripping could help, that is when galaxies, particularly dwarf satellites, pass near the dense cores of filaments, the strong pressure from the surrounding cosmic medium can strip away their gas (\citealt{herzog2023present}).

\subsection{The Whole Sample}\label{sec:compare_observation}

Most previous studies, particularly those based on observations, have analyzed central and satellite galaxies together when examining the relationship between galaxy properties and large-scale environments. However, as demonstrated in the preceding sections, these two populations exhibit significant differences in their environmental dependencies. To facilitate direct comparison with prior research, we present results for the entire galaxy sample in TNG100 in Figure \ref{fig:porperty_together}. The figure reveals pronounced discrepancies between galaxies in groups and those in the other three environments across multiple properties. As discussed earlier, these differences are primarily driven by satellites at the low-mass end and centrals at the high-mass end. Furthermore, noticeable variations in sSFR, gas fraction, color, and age persist between galaxies in transition regions, such as FilDR and GrpDR, and those in the Field, particularly for $\rm{M_\ast} \leq 10^{10.25}\, \rm{M_{\odot}}$. As previously revealed, these differences stem from a combination of local overdensity and cosmic web effects for centrals, while for satellites, they are primarily influenced by host halo mass and local overdensity. 

In our study, the Field environment corresponds to what is referred to as the Void or Field in previous works, while GrpDR and FilDR can be collectively considered as Filaments and Walls in the literature for comparative analysis. The relationships between galaxy properties and the cosmic web environment in TNG100 align well with many previous observational studies. For example, galaxies in low-density voids are reported to be bluer, exhibit higher SFRs, and experience slower star formation histories and stellar evolution compared to those in denser environments (e.g., \citealt{2004ApJ...617...50R,2007ApJ...658..898P,2012MNRAS.426.3041H,2021ApJ...906...97F,dominguez2023galaxies}). Meanwhile, galaxies in filaments and walls appear to be in a transitional phase between those in the Field and those in groups (e.g., \citealt{dominguez2023stellar, zakharova2024virgo}).
Additionally, \cite{dominguez2023stellar} shows that galaxies in filaments and walls exhibit slightly higher, about 0.05 dex, stellar metallicities than those in voids but lower than those in clusters, with differences reaching up to 0.4 dex. These variations are particularly pronounced for low-mass galaxies ($\sim 10^{9.25}\,\mathrm{M_\odot}$). Moreover, galaxies located closer to nodes are found to have higher gas metallicities (\citealt{donnan2022role}).

However, our results diverge moderately from observations in the magnitude of differences in some specific galaxy property seen among diverse environments. For example, the difference in gas fraction between the field and transition regions in our findings is approximately $5\%–10\%$, whereas observations show variations of up to $30\%$ (e.g., \citealt{2021ApJ...906...97F}). Additionally, the differences in metallicity between the group and other regions in our results reach up to 0.1 dex, which is less pronounced than those reported in \cite{dominguez2023stellar}.
This discrepancy may stem primarily from two factors: (1) the limited volume in the TNG100 simulation, which may underestimate the impact of filaments and groups, as it contains fewer massive filaments and clusters compared to observational studies. We will further discuss the impact of the simulation box size and resolution in Section \ref{sec:Compare_results_in_TNG50_TNG100_TNG300}.
(2) difference in redshift range: our results in this Section are based on simulation data at redshift $z = 0.00$, while the results in many previous studies are based on samples with the redshift range $z=0$ and $z\sim 0.5$. This could partially explain why our overall stellar metallicity is somewhat higher.

In summary, our study highlights the impact of the cosmic web environment on galaxies in the TNG100 simulation. After controlling for stellar mass and local overdensity, the influence of cosmic web filaments on central galaxies nearly disappears. However, when stellar mass, host halo mass, and overdensity are controlled, differences in satellite galaxy properties between the field and filament-dominated regions persist, which is very likely driven by physical processes associated with the anisotropic nature of the cosmic web. In contrast, the differences between FilDR and GrpDR almost vanish. These findings highlight the importance of distinguishing between central and satellite galaxies when analyzing the effects of the cosmic web on galaxy evolution.

\subsection{Dependency of galaxy properties on the distance to filaments}\label{sec:property_distance}

\begin{figure*} 
\begin{centering}
\hspace{-0.0cm}
\includegraphics[width=0.9\textwidth]{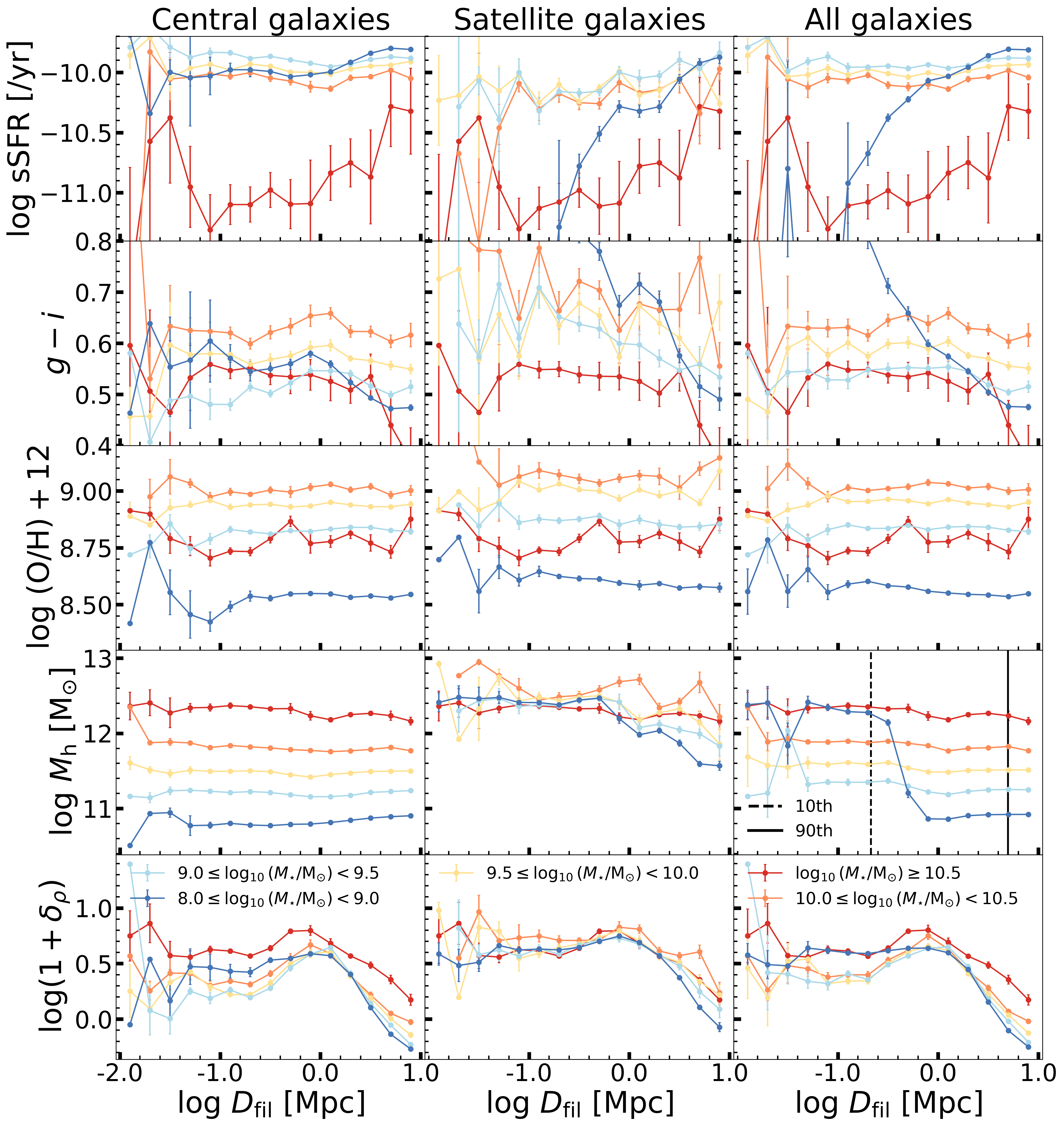}
\caption{The median sSFR, color ($g - i$), gas metallicity, halo mass, overdensity as functions of distance to the nearest filament segment, $D_\mathrm{fil}$, for central (left), satellites (middle) and all galaxies (right). Galaxies in different stellar mass bins are shown with lines of different colors. The lines for galaxies with $\mathit{M}_{\ast} \geq 10^{10.5}\,\rm{M_{\odot}}$ have been shifted upward (downward) by $0.5$ dex in the first (second) row for clarity. In the halo mass panel for all galaxies, the black dashed and solid lines indicate the 10th and 90th percentiles of the $D_\mathrm{fil}$ distribution.}
\label{fig:compare_distribution_distance}
\end{centering}
\end{figure*}

The previous subsections presented results based on our environmental classification. However, many studies in the literature assess the impact of filaments by examining how galaxy properties vary with distance from filaments. To facilitate a direct comparison, we analyze galaxies in the Field, FilDR, and GrpDR regions, investigating their sSFR, color ($g - i$), gas-phase metallicity, halo mass, and local overdensity as functions of distance to the nearest filament segment, $D_\mathrm{fil}$, for central and satellite galaxies and the full galaxy sample. The results are shown in Figure \ref{fig:compare_distribution_distance}.

For central galaxies, we find that within each stellar mass bin, the host halo mass remains relatively constant with respect to $D_\mathrm{fil}$. However, as galaxies move from larger distances toward $D_\mathrm{fil} \sim 1$ Mpc, they tend to have a lower median value of sSFR and redder colors, accompanied by a rapid increase in local overdensity. These trends are more pronounced for galaxies with $M_{\ast} > 10^{10.5}\,\rm{M_{\odot}}$, where the decline in sSFR and shift toward redder colors persist down to $D_\mathrm{fil} \sim 0.3$ Mpc. For less massive galaxies, these trends weaken and largely disappear around $D_\mathrm{fil} \sim 1$ Mpc, after which their properties show little dependence on $D_\mathrm{fil}$ in the range $0.1< D_\mathrm{fil}< 1$ Mpc. Note that local overdensity tends to plateau at $0.1<D_\mathrm{fil}< 1$ Mpc, primarily due to smoothing effects. At $D_\mathrm{fil}<0.1$ Mpc, properties exhibit notable fluctuations due to limited sample sizes, though both sSFR and overdensity tend to increase as $D_\mathrm{fil}$ decreases. Some previous studies have reported similar findings that sSFR could be enhanced in some dense environments at low redshift in both observations (e.g., \citealt{vulcani2019gasp}) and simulations (e.g., \citealt{hasan2023evolving}).

Furthermore, for central galaxies with $10^{9.0}\,\rm{M_{\odot}} < M_{\ast} < 10^{10.5}\,\rm{M_{\odot}}$, we find that sSFR increases and color becomes bluer as $D_\mathrm{fil}$ decreases from 1 Mpc to approximately 0.3 Mpc. This feature is likely driven by a corresponding decrease in local overdensity within this range, resulting in a bump in the overdensity profile within $0.3\, \rm{Mpc}<D_\mathrm{fil}<1.5\, \rm{Mpc}$ for galaxies in this mass range. This bump is likely a result of several contributing factors: (1) the inclusion of galaxies within group-dominated regions (GrpDR), as the bump weakens when these galaxies are excluded; (2) our selection criteria, which exclude short and less massive filament segments, would lead to an overestimation of $D_\mathrm{fil}$ for some galaxies whose true nearest filament segments are either short or less massive; and (3) our definition of filament distance is related to the spine connecting the segment's endpoints. However, for moderately curved filaments, this spine may deviate from the actual density ridge, thus potentially placing galaxies closer to the ridge to locate in the $D_\mathrm{fil}=0.3-1.0$ Mpc range. Consequently, these effects collectively contribute to the observed overdensity bump within the $0.3-1.0$ Mpc in the profile for central galaxies within the specified stellar mass range.

The trends of lower sSFR and redder colors in satellite galaxies are more pronounced than in central galaxies, remaining significant at smaller filament distances, around $0.3,\mathrm{Mpc}$. This is illustrated in the middle column of Figure \ref{fig:compare_distribution_distance} and aligns with previous studies (e.g., \citealt{bulichi2024galaxy}). However, caution is warranted regarding the decline in sSFR for satellites less massive than $10^{9.0}\,\rm{M_{\odot}}$ as TNG100 over produce quenched less massive galaxies. Notably, the halo mass of satellite galaxies increases when they approach the filaments, with $D_\mathrm{fil}$ decreases from $10$ Mpc to 1 Mpc. Moreover, the decrease in color and local overdensity in the range of $0.3<D_\mathrm{fil}<1 $ Mpc for satellite galaxies with $10^{9.0}\,\rm{M_{\odot}} < M_{\ast} < 10^{10.5}\,\rm{M_{\odot}}$ is much weaker compared to the central sample.

When the centrals and satellites are combined, the corresponding trends are shown in the right column of Figure \ref{fig:compare_distribution_distance}. Approximately $10\%$ of the galaxies of reside within $D_\mathrm{fil}<0.25$ Mpc, while $90\%$ are located within $D_\mathrm{fil} < 5$ Mpc (indicated by the black dashed and solid vertical lines in the $M_h-D_\mathrm{fil}$ panel). The evolution of galaxy properties as they approach filaments at $D_\mathrm{fil}>1$ Mpc in TNG aligns qualitatively with findings from both observational and simulation-based studies (e.g., \citealt{kuutma2017voids,kraljic2018galaxy, winkel2021imprint, 2022A&A...658A.113M, bulichi2024galaxy}). However, the slope of these trends exhibits discrepancies. For example, the sSFR of central galaxies in TNG100 decreases by $0.2-0.5$ dex between $D_\mathrm{fil}=10\, \mathrm{Mpc}$ and $\sim 1\, \mathrm{Mpc}$, significantly larger than the $0.04$ dex reported by \cite{2018MNRAS.474..547K}, but comparable to that found in \cite{bulichi2024galaxy}. Meanwhile, Similarly, the color index of centrals in TNG100 increases by $0.1-0.3$ dex, compared to only 0.02 dex in \citealt{kuutma2017voids}.

Several factors may contribute to this discrepancy. First, there are moderate discrepancies between TNG100 and observations on the star formation activity, gas properties and colors. Second, the volume and redshift range of TNG100 are limited, which is subject to cosmic variance. Third, variations in cosmic web filament classification methods across studies, including identify filament central axes and measure of distance to filament, introduce inconsistencies. In addition, moderate variations in the stellar mass ranges considered across studies can also contribute to these discrepancies.

Nevertheless, based on the results presented in previous sections, we argue that assessing the impact of filaments on galaxy properties solely through their dependence on $D_\mathrm{fil}$ is inadequate, as several effects are mixed, including overdensity, halo mass, groups, and the intrinsic influence of the cosmic web.

\begin{figure}[htb]
\begin{center}
\hspace{-0.0cm}
\includegraphics[width=0.45\textwidth]{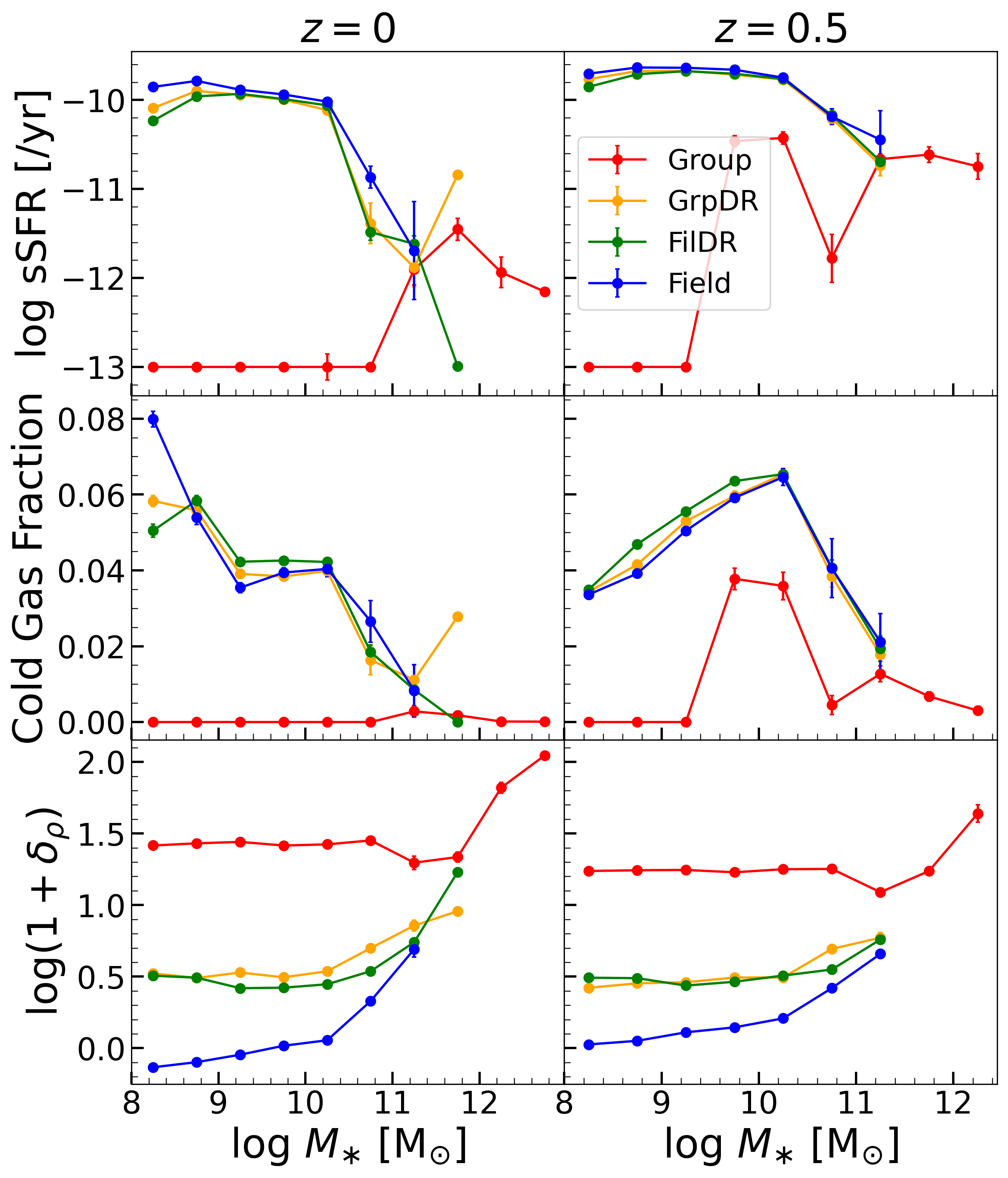}
\caption{The median sSFR and overdensity of whole galaxies in different environments at various redshifts (left: $z = 0$; right: $z = 0.5$) from the TNG100 simulation as a function of stellar mass $M_\ast$.}
\label{fig:compare_redshift_whole}
\end{center}
\end{figure}

\begin{figure*} 
\begin{centering}
\hspace{-0.0cm}
\includegraphics[width=0.75\textwidth]{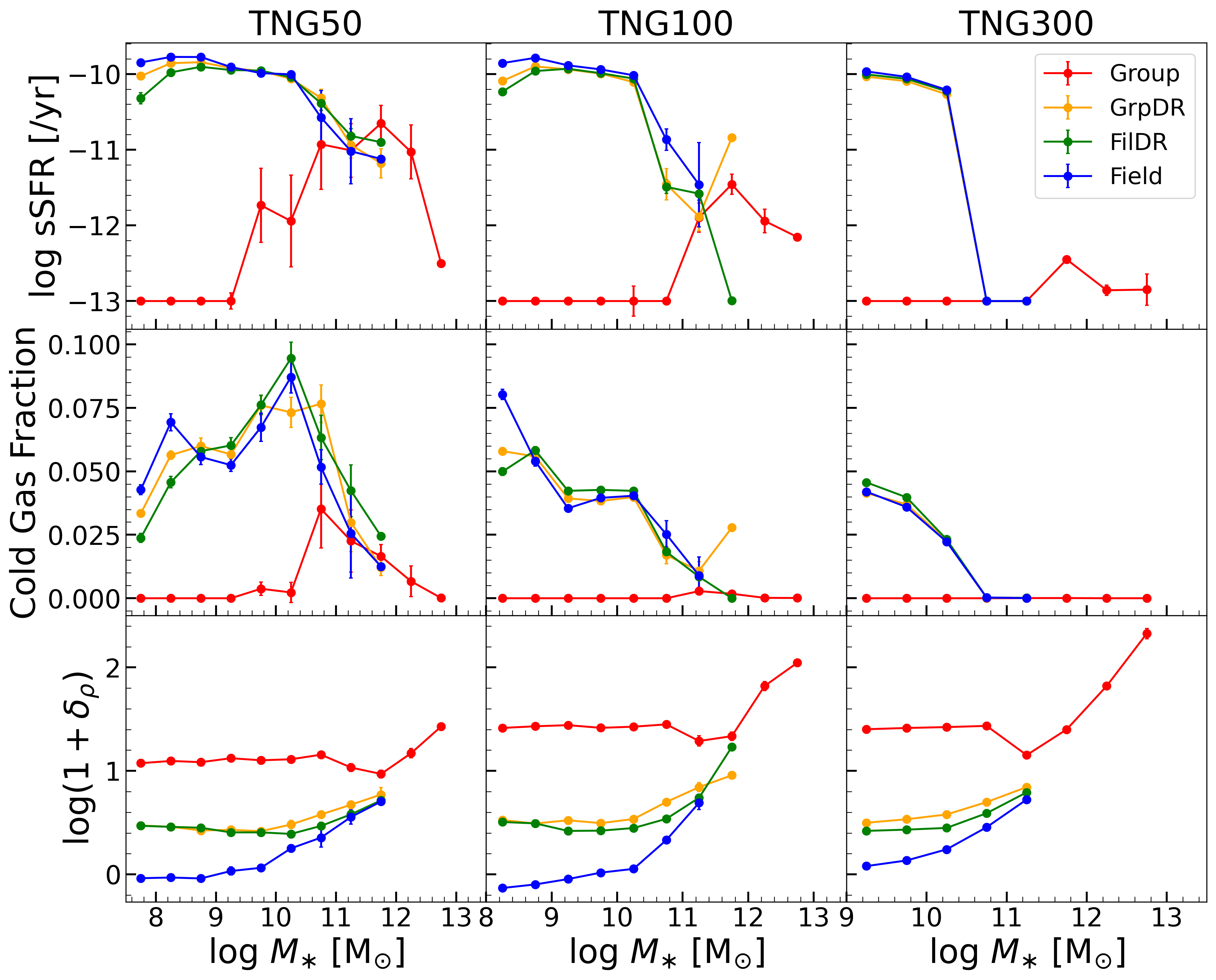}
\caption{The median sSFR and cold gas fraction of whole galaxies in different environment from TNG50 (left), TNG100 (middle), TNG300 (right) simulation as a function of stellar mass $M_\ast$.}
\label{fig:compare_TNG50_100_300_whole}
\end{centering}
\end{figure*}

\section{Discussions}\label{sec:discussions}

\subsection{Results at $z=0.5$}

As large-scale cosmic structures, which emerge from the nonlinear evolution of dark matter, become more prominent at lower redshifts, it is expected that the influence of different environments on galaxy properties would weaken at higher redshifts. To explore this, we examine galaxy samples at redshift $z = 0.5$. For brevity, we focus on the sSFR, cold gas fraction, and overdensity across different environments. The results, shown in Figure \ref{fig:compare_redshift_whole}, clearly demonstrate that the impact of the cosmic web on galaxy properties decreases at $z = 0.5$ for non-group environments.

\subsection{Comparing results in TNG50, TNG100, TNG300}\label{sec:Compare_results_in_TNG50_TNG100_TNG300}

To investigate the effects of simulation resolution and volume, we compare the sSFR, cold gas fraction, and overdensity of the whole galaxies samples from TNG50, TNG100, and TNG300 at $z = 0$ across different environments, as shown in Figure \ref{fig:compare_TNG50_100_300_whole}. The top panel illustrates that for sSFR, galaxies in TNG50 and TNG300 display patterns similar to those in TNG100 for stellar masses below $M_{\ast} < 10^{10.25}\,\mathrm{M_\odot}$. However, at the high-mass end, the results differ significantly, which could be attributed to larger simulation volume containing a greater number of massive galaxies. Additionally, the differences between low-mass galaxies in the Field and those in the transition regions (FilDR and GrpDR) are somewhat weaker in TNG300. The relatively lower resolution in TNG300 may limit its ability to accurately model galaxy properties at the lower mass end.

We also examine properties such as cold gas fraction and overdensity across different simulations. The relationship between the cosmic web environment and cold gas fraction remains ambiguous in all three simulations, likely due to the combined analysis of centrals and satellites. Additionally, there is a clear discrepancy between TNG50 and the other two simulations regarding the overall dependence of the cold gas fraction on stellar mass. In TNG50, the cold gas fraction initially increases with $M_{\ast}$, reaching its peak at $10^{10.25}\,\mathrm{M_\odot}$, before decreasing at higher masses. In contrast, for galaxies in TNG100 and TNG300, the cold gas fraction generally decreases with stellar mass, although with some fluctuations in TNG100. This discrepancy is likely caused by the following reasons. The higher resolution in TNG50 can enhance the efficiency of ISM heating by stellar feedback, and resolve small-scale cool clouds instead of a more homogeneus cold phase of gas (\citealt{nelson2020resolving}). Moreover, the cold gas content is not converged event with the TNG50 resolution (\citealt{peeples2019figuring,mandelker2021thermal}). Higher resolution is needed to accurately assess the impact of cosmic web on cold gas in galaxies. 

On the other hand, the relationship between overdensity and stellar mass across different environments remains consistent tendency across all three simulations. However, group galaxies in TNG50 have lower overdensity than that TNG100 or TNG300, because TNG50's small volume lead to lack of massive clusters with respect to the other two simulations.

Overall, the influence of different cosmic web environments on galaxy properties exhibits moderate discrepancies across the three simulations. To achieve a more comprehensive evaluation, a simulation with a volume comparable to TNG300 and a resolution similar to TNG50 or TNG100 would be highly beneficial.

\begin{figure}[htb]
\begin{center}
\hspace{-0.0cm}
\includegraphics[width=0.45\textwidth]{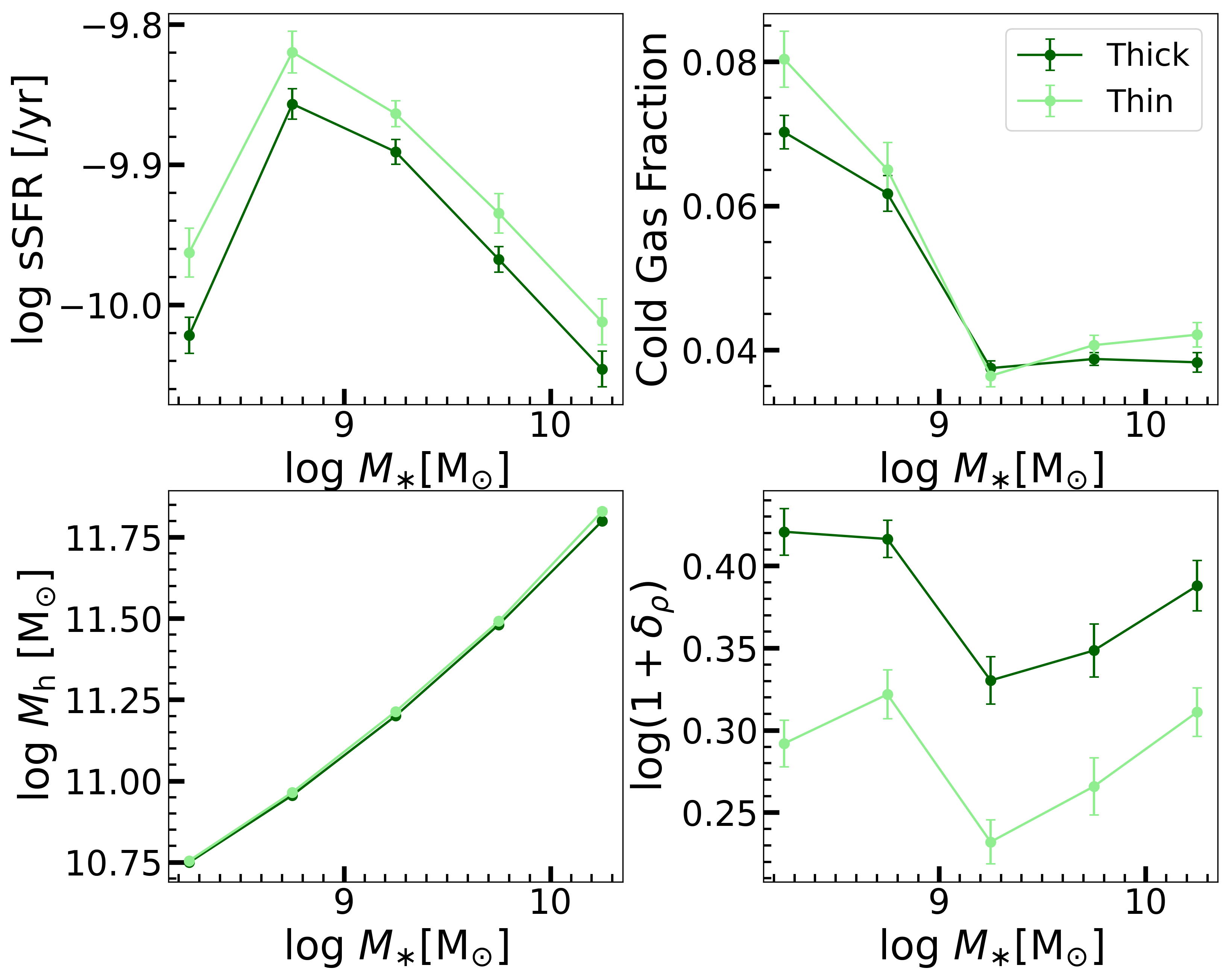}
\caption{The median sSFR and cold gas fraction, halo mass and overdensity for central galaxies with $ M_{\ast} <= 10^{10.25}\, \mathrm{M_\odot}$ in thick filaments with radius $R_{\mathrm{fil}} \geq 2\, \mathrm{Mpc}$, and thin filaments as a function of stellar mass \( M_\ast \).  Light and dark green lines correspond to thin and thick filaments, respectively. There are 3,328 and 6,312 galaxies in thick and thin filaments respectively.}
\label{fig:thickness_central}
\end{center}
\end{figure}

\subsection{Impact of filaments with various thickness}

Cosmic filaments exhibit distinct variations in length and thickness (e.g., \citealt{2014MNRAS.441.2923C,2021ApJ...920....2Z}). \cite{2022ApJ...924..132Z} demonstrated that the gas accretion rate onto low-mass halos is lower in thick filaments compared to thin filaments. This suggests that filament thickness may influence galaxy properties in different ways. To investigate this potential effect, we divided the central galaxy sample in FilDR from TNG100 into two sub-samples: those located near thick filaments ($R_{\mathrm{fil}} \geq 2 \mathrm{Mpc}$) and those near thin filaments ($R_{\mathrm{fil}} < 2 \mathrm{Mpc}$). Figure \ref{fig:thickness_central} illustrates the properties of central galaxies in these two sub-samples, including sSFR, cold gas fraction, $M_\mathrm{h}$, and overdensity. 

For low-mass galaxies ($M_{\ast} < 10^{9.5}\,\mathrm{M_\odot}$), those located near thin filaments exhibit slightly to moderately higher sSFR, greater cold gas fractions, and lower overdensity compared to their counterparts near thick filaments. Since the halo mass for galaxies of a given stellar mass is similar between these two sub-samples, the observed differences in sSFR and cold gas fraction are likely influenced, at least in part, by the surrounding overdensity. Additionally, thick filaments are known to contain hotter gas (\citealt{2021ApJ...920....2Z,2022ApJ...924..132Z}; Yang et al., in prep.), which may further inhibit the accretion of cold gas and suppress star formation activity. 

Consistent with our findings, \cite{hasan2023filaments} demonstrated using TNG100 simulations that galaxies residing near denser filaments exhibit reduced gas fractions and star formation rates at $z \lesssim 1$, compared to those near less dense filaments. Observationally, \cite{odekon2018effect} reported $M_\mathrm{H\,\textsc{i}}$-deficiencies in galaxies within thin filaments ("tendrils"), suggesting a closer resemblance to void galaxies than to typical thick-filament populations. Similarly, \cite{alpaslan2016galaxy} observed a reduction in sSFR within these tendrils. These results are likely indicative of significant physical processes that regulate gas transfer from filaments to galaxies or halos. However, a more detailed and comprehensive analysis is required to establish definitive conclusions regarding this aspect. Furthermore, the development of robust methods to accurately measure filament widths is crucial (\citealt{2024ApJ...967..141Z}).

\subsection{Limitations}
There are certain limitations in our study that may lead to an underestimation of the impact of the cosmic web environment on galaxy properties. First, while the IllustrisTNG simulations broadly align with observational trends across various galaxy properties, they still exhibit notable discrepancies. For instance, the simulated galaxy stellar mass function (GSMF) is slightly to moderately higher than the observational estimates at both the low- and high-mass ends. Additionally, the simulation produces a moderately higher fraction of quenched dwarf galaxies, red discs and blue spheroids compared to observations (\citealt{2021MNRAS.502.1051A,crain2023hydrodynamical}). Also, TNG has less of a green valley than observed at low redshifts, very likely due to the rapid quenching driven by the kinetic AGN feedback (\citealt{donnari2021quenched}).

Moreover, IllustrisTNG overestimates the total neutral gas abundance at $z=0$ by approximately a factor of two, predicts an excess of galaxies with $M_\mathrm{H\,\textsc{i}} \sim 10^9\,\mathrm{M_\odot}$, and generates an overabundance of satellite galaxies that contain little or no neutral gas (\citealt{2019MNRAS.487.1529D}). TNG also slightly overproduces gas fractions in massive galaxies relative to observations (\citealt{stevens2019origin}).
These discrepancies between the galaxy population in TNG100 and those analyzed in previous observational and simulation studies likely contribute to the partial inconsistencies between our results and findings in the literature.

Second, several galaxy properties in IllustrisTNG exhibit a degree of resolution dependence due to the implementation of sub-grid physics and the choice of model parameters (e.g., \citealt{2018MNRAS.473.4077P,crain2023hydrodynamical}). Furthermore, quantitative discrepancies in galaxy properties between TNG and other simulations persist, stemming from variations in sub-grid physics model implementations. For instance, differing active galactic nuclei and supernova feedback prescriptions result in a substantially higher fraction of $M_\mathrm{H\,\textsc{i}}$ gas in low-mass galaxies within the TNG simulation compared to EAGLE (\citealt{dave2020galaxy,li2024drives}).

Third, the limited volume of TNG100 may lead to an under representation of prominent large-scale structures, potentially underestimating their influence on galaxy properties. Also, the classification methods used to define the cosmic web vary significantly across different studies (e.g., \citealt{chen2017detecting,2019MNRAS.483.3227K,kuchner2020mapping,kuchner2021cosmic,hasan2023filaments,hasan2023evolving}). While our classification approach aligns with methodologies adopted in several previous works (e.g., \citealt{2019MNRAS.483.3227K,song2021beyond,bulichi2024galaxy,o2024effect}), variations in the specific procedures used can introduce discrepancies. 

Additionally, variations in the stellar mass threshold used to define cosmic web tracer galaxies can influence filament identification. Specifically, lowering the threshold (e.g., from $ 10^{9}\,\mathrm{M_\odot}$ down to $ 10^{8.5}\,\mathrm{M_\odot}$), may resolve thick filaments into multiple tenuous structures, while raising the threshold would exclude a fraction of these tenuous filaments. We anticipate that these differences in cosmic web classification introduce minor quantitative variations in our results, potentially contributing to the observed discrepancies between our findings and those reported in the literature. 


\section{Conclusions}\label{sec:conclusions}

In this study, we explore the influence of the cosmic web on galaxy properties using the IllustrisTNG simulations. To disentangle the effects of groups and clusters from those of filaments, we introduce an environmental classification method that categorizes the cosmic web into four distinct regions: Group, GrpDR (group-dominated regions), FilDR (filament-dominated regions), and Field. We systematically compare the properties of galaxies with the same stellar mass across different cosmic web environments, analyzing central and satellite galaxies separately. To isolate the impact of cosmic web filaments, we further control for the effects of overdensity and host dark matter halo mass. Our key findings can be summarized as follows.

\begin{enumerate}
    \item In IllustrisTNG, central galaxies of similar stellar and halo mass exhibit obvious differences in their properties across different cosmic web environments, particularly for low-mass galaxies. Field galaxies tend to exhibit higher sSFR and gas fractions, lower local densities, and younger stellar populations compared to those in FilDR and GrpDR, following an ordered trend. These variations are primarily driven by local overdensity. After controlling for both stellar mass and local overdensity, the influence of cosmic web filaments on central galaxies nearly disappears.
    
    \item Satellite galaxies within IllustrisTNG, when matched in stellar mass, display significant property variations across different cosmic web environments. Although halo mass and local overdensity are primary drivers of these variations, residual differences between satellites residing in field and filament-dominated regions remain even after controlling for these factors. This suggests that the cosmic web environment itself exerts an independent influence on satellite galaxy properties. Notably, satellite galaxies appear more affected by the cosmic web than central galaxies. The distinct physical processes governing central and satellite galaxy evolution necessitate their separate analysis for accurate assessment of environmental effects.

    \item As galaxies approaching filaments in IllustrisTNG, move from $D_\mathrm{fil} \sim 10$ Mpc to $D_\mathrm{fil} \sim 1$ Mpc, they generally exhibit a $0.2-0.5$ dex decrease in specific star formation rate (sSFR) and a shift toward redder colors by $0.1-0.3$ dex. However, the relationships between galaxy properties and their distance from filaments arise from a combination of factors, including stellar mass, halo mass, groups, overdensity, and the intrinsic influence of the cosmic web on galaxy evolution.
    
    \item The impact of the cosmic web on galaxy properties at redshift $z=0.5$ is weaker than at $z=0$. Additionally, thick filaments are often accompanied by higher overdensity. Central galaxies near thick filaments tend to exhibit slightly or moderately lower sSFR and cold gas fractions compared to those near thin filaments.
    
\end{enumerate}

The trends we report, namely that galaxies with the same stellar mass in the TNG simulation exhibit different properties across various cosmic environments, with those in denser regions tending to have lower specific star formation rates (sSFR), lower gas fractions, and older stellar populations, are consistent with observational studies(\citealt{2004ApJ...617...50R,2007ApJ...658..898P,2012MNRAS.426.3041H,2021ApJ...906...97F,dominguez2023galaxies,zakharova2024virgo}). Additionally, the trend in TNG, where galaxies closer to filaments show lower sSFR and redder colors, aligns with observational results (\citealt{alpaslan2016galaxy, chen2017detecting, kraljic2018galaxy, 2018MNRAS.474.5437L, winkel2021imprint, castignani2022virgo, laigle2025euclid}).

However, the rate at which galaxy properties change across different environments shows a noticeable discrepancy between TNG and observations. Additionally, another discrepancy exists in the slope of these property variations with filament distance. These differences likely stem from several factors, including intrinsic variations in galaxy properties between TNG and real observations, differences in cosmic web classification methods, the limited volume of TNG100, and variations in stellar mass and redshift ranges across datasets.

Our findings highlight the importance of distinguishing between central and satellite galaxies to accurately evaluate the environmental effects of the cosmic web. Additionally, controlling for stellar mass, halo mass, and local overdensity is essential to isolate the intrinsic influence of the cosmic web on galaxy evolution.

\acknowledgments
We thank the anonymous referee for providing very useful comments and suggestions that improved the manuscript. We acknowledge the helpful discussions with Antonios Katsianis, Xichang Ouyang, Junyu Gong, and Cheng Li. This work is supported by the National Natural Science Foundation of China (NFSC) through grant 11733010. W.S.Z. is supported by NSFC grant 12173102. The calculation carried out in this work was completed on the HPC facility of the School of Physics and Astronomy, Sun, Yat-Sen University.

%




\vspace{15mm}





\bibliography{main}{}

\begin{thebibliography}{}
\expandafter\ifx\csname natexlab\endcsname\relax\def\natexlab#1{#1}\fi
\providecommand{\url}[1]{\href{#1}{#1}}
\providecommand{\dodoi}[1]{doi:~\href{http://doi.org/#1}{\nolinkurl{#1}}}
\providecommand{\doeprint}[1]{\href{http://ascl.net/#1}{\nolinkurl{http://ascl.net/#1}}}
\providecommand{\doarXiv}[1]{\href{https://arxiv.org/abs/#1}{\nolinkurl{https://arxiv.org/abs/#1}}}

\bibitem[{{Alpaslan} {et~al.}(2014){Alpaslan}, {Robotham}, {Driver}, {Norberg}, {Baldry}, {Bauer}, {Bland-Hawthorn}, {Brown}, {Cluver}, {Colless}, {Foster}, {Hopkins}, {Van Kampen}, {Kelvin}, {Lara-Lopez}, {Liske}, {Lopez-Sanchez}, {Loveday}, {McNaught-Roberts}, {Merson}, \& {Pimbblet}}]{2014MNRAS.438..177A}
{Alpaslan}, M., {Robotham}, A. S.~G., {Driver}, S., {et~al.} 2014, \mnras, 438, 177, \dodoi{10.1093/mnras/stt2136}

\bibitem[{{Alpaslan} {et~al.}(2015){Alpaslan}, {Driver}, {Robotham}, {Obreschkow}, {Andrae}, {Cluver}, {Kelvin}, {Lange}, {Owers}, {Taylor}, {Andrews}, {Bamford}, {Bland-Hawthorn}, {Brough}, {Brown}, {Colless}, {Davies}, {Eardley}, {Grootes}, {Hopkins}, {Kennedy}, {Liske}, {Lara-L{\'o}pez}, {L{\'o}pez-S{\'a}nchez}, {Loveday}, {Madore}, {Mahajan}, {Meyer}, {Moffett}, {Norberg}, {Penny}, {Pimbblet}, {Popescu}, {Seibert}, \& {Tuffs}}]{alpaslan2015galaxy}
{Alpaslan}, M., {Driver}, S., {Robotham}, A. S.~G., {et~al.} 2015, \mnras, 451, 3249, \dodoi{10.1093/mnras/stv1176}

\bibitem[{{Alpaslan} {et~al.}(2016){Alpaslan}, {Grootes}, {Marcum}, {Popescu}, {Tuffs}, {Bland-Hawthorn}, {Brough}, {Brown}, {Davies}, {Driver}, {Holwerda}, {Kelvin}, {Lara-L{\'o}pez}, {L{\'o}pez-S{\'a}nchez}, {Loveday}, {Moffett}, {Taylor}, {Owers}, \& {Robotham}}]{alpaslan2016galaxy}
{Alpaslan}, M., {Grootes}, M., {Marcum}, P.~M., {et~al.} 2016, \mnras, 457, 2287, \dodoi{10.1093/mnras/stw134}

\bibitem[{{Arag{\'o}n-Calvo} {et~al.}(2007){Arag{\'o}n-Calvo}, {Jones}, {van de Weygaert}, \& {van der Hulst}}]{2007A&A...474..315A}
{Arag{\'o}n-Calvo}, M.~A., {Jones}, B.~J.~T., {van de Weygaert}, R., \& {van der Hulst}, J.~M. 2007, \aap, 474, 315, \dodoi{10.1051/0004-6361:20077880}

\bibitem[{{Arag{\'o}n-Calvo} {et~al.}(2010{\natexlab{a}}){Arag{\'o}n-Calvo}, {Platen}, {van de Weygaert}, \& {Szalay}}]{2010ApJ...723..364A}
{Arag{\'o}n-Calvo}, M.~A., {Platen}, E., {van de Weygaert}, R., \& {Szalay}, A.~S. 2010{\natexlab{a}}, \apj, 723, 364, \dodoi{10.1088/0004-637X/723/1/364}

\bibitem[{{Arag{\'o}n-Calvo} {et~al.}(2010{\natexlab{b}}){Arag{\'o}n-Calvo}, {van de Weygaert}, \& {Jones}}]{2010MNRAS.408.2163A}
{Arag{\'o}n-Calvo}, M.~A., {van de Weygaert}, R., \& {Jones}, B. J.~T. 2010{\natexlab{b}}, \mnras, 408, 2163, \dodoi{10.1111/j.1365-2966.2010.17263.x}

\bibitem[{{Ayromlou} {et~al.}(2021){Ayromlou}, {Nelson}, {Yates}, {Kauffmann}, {Renneby}, \& {White}}]{2021MNRAS.502.1051A}
{Ayromlou}, M., {Nelson}, D., {Yates}, R.~M., {et~al.} 2021, \mnras, 502, 1051, \dodoi{10.1093/mnras/staa4011}

\bibitem[{{Bond} {et~al.}(1996){Bond}, {Kofman}, \& {Pogosyan}}]{1996Natur.380..603B}
{Bond}, J.~R., {Kofman}, L., \& {Pogosyan}, D. 1996, \nat, 380, 603, \dodoi{10.1038/380603a0}

\bibitem[{{Bulichi} {et~al.}(2024){Bulichi}, {Dav{\'e}}, \& {Kraljic}}]{bulichi2024galaxy}
{Bulichi}, T.-E., {Dav{\'e}}, R., \& {Kraljic}, K. 2024, \mnras, 529, 2595, \dodoi{10.1093/mnras/stae667}

\bibitem[{{Castignani} {et~al.}(2022){Castignani}, {Vulcani}, {Finn}, {Combes}, {Jablonka}, {Rudnick}, {Zaritsky}, {Whalen}, {Conger}, {De Lucia}, {Desai}, {Koopmann}, {Moustakas}, {Norman}, \& {Townsend}}]{castignani2022virgo}
{Castignani}, G., {Vulcani}, B., {Finn}, R.~A., {et~al.} 2022, \apjs, 259, 43, \dodoi{10.3847/1538-4365/ac45f7}

\bibitem[{{Cautun} {et~al.}(2014){Cautun}, {van de Weygaert}, {Jones}, \& {Frenk}}]{2014MNRAS.441.2923C}
{Cautun}, M., {van de Weygaert}, R., {Jones}, B. J.~T., \& {Frenk}, C.~S. 2014, \mnras, 441, 2923, \dodoi{10.1093/mnras/stu768}

\bibitem[{{Chen} {et~al.}(2017){Chen}, {Ho}, {Mandelbaum}, {Bahcall}, {Brownstein}, {Freeman}, {Genovese}, {Schneider}, \& {Wasserman}}]{chen2017detecting}
{Chen}, Y.-C., {Ho}, S., {Mandelbaum}, R., {et~al.} 2017, \mnras, 466, 1880, \dodoi{10.1093/mnras/stw3127}

\bibitem[{{Colberg} {et~al.}(2005){Colberg}, {Krughoff}, \& {Connolly}}]{2005MNRAS.359..272C}
{Colberg}, J.~M., {Krughoff}, K.~S., \& {Connolly}, A.~J. 2005, \mnras, 359, 272, \dodoi{10.1111/j.1365-2966.2005.08897.x}

\bibitem[{{Colless} {et~al.}(2003){Colless}, {Peterson}, {Jackson}, {Peacock}, {Cole}, {Norberg}, {Baldry}, {Baugh}, {Bland-Hawthorn}, {Bridges}, {Cannon}, {Collins}, {Couch}, {Cross}, {Dalton}, {De Propris}, {Driver}, {Efstathiou}, {Ellis}, {Frenk}, {Glazebrook}, {Lahav}, {Lewis}, {Lumsden}, {Maddox}, {Madgwick}, {Sutherland}, \& {Taylor}}]{2003astro.ph..6581C}
{Colless}, M., {Peterson}, B.~A., {Jackson}, C., {et~al.} 2003, arXiv e-prints, astro, \dodoi{10.48550/arXiv.astro-ph/0306581}

\bibitem[{{Crain} \& {van de Voort}(2023)}]{crain2023hydrodynamical}
{Crain}, R.~A., \& {van de Voort}, F. 2023, \araa, 61, 473, \dodoi{10.1146/annurev-astro-041923-043618}

\bibitem[{{Crone Odekon} {et~al.}(2018){Crone Odekon}, {Hallenbeck}, {Haynes}, {Koopmann}, {Phi}, \& {Wolfe}}]{odekon2018effect}
{Crone Odekon}, M., {Hallenbeck}, G., {Haynes}, M.~P., {et~al.} 2018, \apj, 852, 142, \dodoi{10.3847/1538-4357/aaa1e8}

\bibitem[{{Darvish} {et~al.}(2016){Darvish}, {Mobasher}, {Sobral}, {Rettura}, {Scoville}, {Faisst}, \& {Capak}}]{darvish2016effects}
{Darvish}, B., {Mobasher}, B., {Sobral}, D., {et~al.} 2016, \apj, 825, 113, \dodoi{10.3847/0004-637X/825/2/113}

\bibitem[{{Dav{\'e}} {et~al.}(2019){Dav{\'e}}, {Angl{\'e}s-Alc{\'a}zar}, {Narayanan}, {Li}, {Rafieferantsoa}, \& {Appleby}}]{2019MNRAS.486.2827D}
{Dav{\'e}}, R., {Angl{\'e}s-Alc{\'a}zar}, D., {Narayanan}, D., {et~al.} 2019, \mnras, 486, 2827, \dodoi{10.1093/mnras/stz937}

\bibitem[{{Dav{\'e}} {et~al.}(2020){Dav{\'e}}, {Crain}, {Stevens}, {Narayanan}, {Saintonge}, {Catinella}, \& {Cortese}}]{dave2020galaxy}
{Dav{\'e}}, R., {Crain}, R.~A., {Stevens}, A. R.~H., {et~al.} 2020, \mnras, 497, 146, \dodoi{10.1093/mnras/staa1894}

\bibitem[{{de Lapparent} {et~al.}(1986){de Lapparent}, {Geller}, \& {Huchra}}]{1986ApJ...302L...1D}
{de Lapparent}, V., {Geller}, M.~J., \& {Huchra}, J.~P. 1986, \apjl, 302, L1, \dodoi{10.1086/184625}

\bibitem[{{Diemer} {et~al.}(2019){Diemer}, {Stevens}, {Lagos}, {Calette}, {Tacchella}, {Hernquist}, {Marinacci}, {Nelson}, {Pillepich}, {Rodriguez-Gomez}, {Villaescusa-Navarro}, \& {Vogelsberger}}]{2019MNRAS.487.1529D}
{Diemer}, B., {Stevens}, A. R.~H., {Lagos}, C. d.~P., {et~al.} 2019, \mnras, 487, 1529, \dodoi{10.1093/mnras/stz1323}

\bibitem[{{Dom{\'\i}nguez-G{\'o}mez} {et~al.}(2023{\natexlab{a}}){Dom{\'\i}nguez-G{\'o}mez}, {P{\'e}rez}, {Ruiz-Lara}, {Peletier}, {S{\'a}nchez-Bl{\'a}zquez}, {Lisenfeld}, {Falc{\'o}n-Barroso}, {Alc{\'a}zar-Laynez}, {Argudo-Fern{\'a}ndez}, {Bl{\'a}zquez-Calero}, {Courtois}, {Duarte Puertas}, {Espada}, {Florido}, {Garc{\'\i}a-Benito}, {Jim{\'e}nez}, {Kreckel}, {Rela{\~n}o}, {S{\'a}nchez-Menguiano}, {van der Hulst}, {van de Weygaert}, {Verley}, \& {Zurita}}]{dominguez2023galaxies}
{Dom{\'\i}nguez-G{\'o}mez}, J., {P{\'e}rez}, I., {Ruiz-Lara}, T., {et~al.} 2023{\natexlab{a}}, \nat, 619, 269, \dodoi{10.1038/s41586-023-06109-1}

\bibitem[{{Dom{\'\i}nguez-G{\'o}mez} {et~al.}(2023{\natexlab{b}}){Dom{\'\i}nguez-G{\'o}mez}, {P{\'e}rez}, {Ruiz-Lara}, {Peletier}, {S{\'a}nchez-Bl{\'a}zquez}, {Lisenfeld}, {Bidaran}, {Falc{\'o}n-Barroso}, {Alc{\'a}zar-Laynez}, {Argudo-Fern{\'a}ndez}, {Bl{\'a}zquez-Calero}, {Courtois}, {Duarte Puertas}, {Espada}, {Florido}, {Garc{\'\i}a-Benito}, {Jim{\'e}nez}, {Kreckel}, {Rela{\~n}o}, {S{\'a}nchez-Menguiano}, {van der Hulst}, {van de Weygaert}, {Verley}, \& {Zurita}}]{dominguez2023stellar}
---. 2023{\natexlab{b}}, \aap, 680, A111, \dodoi{10.1051/0004-6361/202346884}

\bibitem[{{Donnan} {et~al.}(2022){Donnan}, {Tojeiro}, \& {Kraljic}}]{donnan2022role}
{Donnan}, C.~T., {Tojeiro}, R., \& {Kraljic}, K. 2022, Nature Astronomy, 6, 599, \dodoi{10.1038/s41550-022-01619-w}

\bibitem[{{Donnari} {et~al.}(2021){Donnari}, {Pillepich}, {Joshi}, {Nelson}, {Genel}, {Marinacci}, {Rodriguez-Gomez}, {Pakmor}, {Torrey}, {Vogelsberger}, \& {Hernquist}}]{donnari2021quenched}
{Donnari}, M., {Pillepich}, A., {Joshi}, G.~D., {et~al.} 2021, \mnras, 500, 4004, \dodoi{10.1093/mnras/staa3006}

\bibitem[{{Dressler}(1980)}]{1980ApJ...236..351D}
{Dressler}, A. 1980, \apj, 236, 351, \dodoi{10.1086/157753}

\bibitem[{{Dubois} {et~al.}(2014){Dubois}, {Pichon}, {Welker}, {Le Borgne}, {Devriendt}, {Laigle}, {Codis}, {Pogosyan}, {Arnouts}, {Benabed}, {Bertin}, {Blaizot}, {Bouchet}, {Cardoso}, {Colombi}, {de Lapparent}, {Desjacques}, {Gavazzi}, {Kassin}, {Kimm}, {McCracken}, {Milliard}, {Peirani}, {Prunet}, {Rouberol}, {Silk}, {Slyz}, {Sousbie}, {Teyssier}, {Tresse}, {Treyer}, {Vibert}, \& {Volonteri}}]{2014MNRAS.444.1453D}
{Dubois}, Y., {Pichon}, C., {Welker}, C., {et~al.} 2014, \mnras, 444, 1453, \dodoi{10.1093/mnras/stu1227}

\bibitem[{Efron(1992)}]{efron1992bootstrap}
Efron, B. 1992, Bootstrap Methods: Another Look at the Jackknife, ed. S.~Kotz \& N.~L. Johnson (New York, NY: Springer New York), 569--593, \dodoi{10.1007/978-1-4612-4380-9_41}

\bibitem[{{Euclid Collaboration} {et~al.}(2025){Euclid Collaboration}, {Laigle}, {Gouin}, {Sarron}, {Quilley}, {Pichon}, {Kraljic}, {Durret}, {Chisari}, {Kuchner}, {Malavasi}, {Magliocchetti}, {McCracken}, {Sorce}, {Kang}, {McPartland}, {Toft}, {Aghanim}, {Altieri}, {Amara}, {Andreon}, {Auricchio}, {Aussel}, {Baccigalupi}, {Baldi}, {Balestra}, {Bardelli}, {Basset}, {Battaglia}, {Bernardeau}, {Biviano}, {Bonchi}, {Branchini}, {Brescia}, {Brinchmann}, {Camera}, {Ca{\~n}as-Herrera}, {Capobianco}, {Carbone}, {Carretero}, {Casas}, {Castellano}, {Castignani}, {Cavuoti}, {Chambers}, {Cimatti}, {Colodro-Conde}, {Congedo}, {Conselice}, {Conversi}, {Copin}, {Courbin}, {Courtois}, {Cropper}, {Da Silva}, {Degaudenzi}, {de la Torre}, {De Lucia}, {Di Giorgio}, {Dolding}, {Dole}, {Dubath}, {Duncan}, {Dupac}, {Ealet}, {Escoffier}, {Farina}, {Farinelli}, {Faustini}, {Ferriol}, {Finelli}, {Fotopoulou}, {Frailis}, {Franceschi}, {Galeotta}, {George}, {Gillard}, {Gillis}, {Giocoli}, {G{\'o}mez-Alvarez}, {Gracia-Carpio},
  {Granett}, {Grazian}, {Grupp}, {Gwyn}, {Haugan}, {Hoekstra}, {Holmes}, {Hook}, {Hormuth}, {Hornstrup}, {Hudelot}, {Jahnke}, {Jhabvala}, {Joachimi}, {Keih{\"a}nen}, {Kermiche}, {Kiessling}, {Kilbinger}, {Kubik}, {Kuijken}, {K{\"u}mmel}, {Kunz}, {Kurki-Suonio}, {Le Boulc'h}, {Le Brun}, {Le Mignant}, {Liebing}, {Ligori}, {Lilje}, {Lindholm}, {Lloro}, {Mainetti}, {Maino}, {Maiorano}, {Mansutti}, {Marcin}, {Marggraf}, {Martinelli}, {Martinet}, {Marulli}, {Massey}, {Maurogordato}, {Medinaceli}, {Mei}, {Melchior}, {Mellier}, {Meneghetti}, {Merlin}, {Meylan}, {Mora}, {Moresco}, {Moscardini}, {Nakajima}, {Neissner}, {Niemi}, {Nightingale}, {Padilla}, {Paltani}, {Pasian}, {Pedersen}, {Percival}, {Pettorino}, {Pires}, {Polenta}, {Poncet}, {Popa}, {Pozzetti}, {Raison}, {Rebolo}, {Renzi}, {Rhodes}, {Riccio}, {Romelli}, {Roncarelli}, {Rusholme}, {Saglia}, {Sakr}, {S{\'a}nchez}, {Sapone}, {Sartoris}, {Schewtschenko}, {Schirmer}, {Schneider}, {Schrabback}, {Scodeggio}, {Secroun}, {Seidel}, {Seiffert}, {Serrano}, {Simon},
  {Sirignano}, {Sirri}, {Skottfelt}, {Stanco}, {Steinwagner}, {Tallada-Cresp{\'\i}}, {Taylor}, {Teplitz}, {Tereno}, {Tessore}, {Toledo-Moreo}, {Torradeflot}, {Tutusaus}, {Valenziano}, {Valiviita}, {Vassallo}, {Verdoes Kleijn}, {Veropalumbo}, {Wang}, {Weller}, {Zacchei}, {Zamorani}, {Zerbi}, {Zinchenko}, {Zucca}, {Allevato}, {Ballardini}, {Bolzonella}, \& {Bozzo}}]{laigle2025euclid}
{Euclid Collaboration}, {Laigle}, C., {Gouin}, C., {et~al.} 2025, arXiv e-prints, arXiv:2503.15333, \dodoi{10.48550/arXiv.2503.15333}

\bibitem[{{Florez} {et~al.}(2021){Florez}, {Berlind}, {Kannappan}, {Stark}, {Eckert}, {Calderon}, {Moffett}, {Campbell}, \& {Sinha}}]{2021ApJ...906...97F}
{Florez}, J., {Berlind}, A.~A., {Kannappan}, S.~J., {et~al.} 2021, \apj, 906, 97, \dodoi{10.3847/1538-4357/abca9f}

\bibitem[{{Forero-Romero} {et~al.}(2009){Forero-Romero}, {Hoffman}, {Gottl{\"o}ber}, {Klypin}, \& {Yepes}}]{2009MNRAS.396.1815F}
{Forero-Romero}, J.~E., {Hoffman}, Y., {Gottl{\"o}ber}, S., {Klypin}, A., \& {Yepes}, G. 2009, \mnras, 396, 1815, \dodoi{10.1111/j.1365-2966.2009.14885.x}

\bibitem[{{Gabor} \& {Dav{\'e}}(2015)}]{2015MNRAS.447..374G}
{Gabor}, J.~M., \& {Dav{\'e}}, R. 2015, \mnras, 447, 374, \dodoi{10.1093/mnras/stu2399}

\bibitem[{{Gunn} \& {Gott}(1972)}]{gunn1972infall}
{Gunn}, J.~E., \& {Gott}, III, J.~R. 1972, \apj, 176, 1, \dodoi{10.1086/151605}

\bibitem[{{Hasan} {et~al.}(2023){Hasan}, {Burchett}, {Abeyta}, {Hellinger}, {Mandelker}, {Primack}, {Faber}, {Koo}, {Elek}, \& {Nagai}}]{hasan2023evolving}
{Hasan}, F., {Burchett}, J.~N., {Abeyta}, A., {et~al.} 2023, \apj, 950, 114, \dodoi{10.3847/1538-4357/acd11c}

\bibitem[{{Hasan} {et~al.}(2024){Hasan}, {Burchett}, {Hellinger}, {Elek}, {Nagai}, {Faber}, {Primack}, {Koo}, {Mandelker}, \& {Woo}}]{hasan2023filaments}
{Hasan}, F., {Burchett}, J.~N., {Hellinger}, D., {et~al.} 2024, \apj, 970, 177, \dodoi{10.3847/1538-4357/ad4ee2}

\bibitem[{{Herzog} {et~al.}(2023){Herzog}, {Ben{\'\i}tez-Llambay}, \& {Fumagalli}}]{herzog2023present}
{Herzog}, G., {Ben{\'\i}tez-Llambay}, A., \& {Fumagalli}, M. 2023, \mnras, 518, 6305, \dodoi{10.1093/mnras/stac3282}

\bibitem[{{Hoffman} {et~al.}(2012){Hoffman}, {Metuki}, {Yepes}, {Gottl{\"o}ber}, {Forero-Romero}, {Libeskind}, \& {Knebe}}]{2012MNRAS.425.2049H}
{Hoffman}, Y., {Metuki}, O., {Yepes}, G., {et~al.} 2012, \mnras, 425, 2049, \dodoi{10.1111/j.1365-2966.2012.21553.x}

\bibitem[{{Hoosain} {et~al.}(2024){Hoosain}, {Blyth}, {Skelton}, {Kannappan}, {Stark}, {Eckert}, {Hutchens}, {Carr}, \& {Kraljic}}]{hoosain2024effect}
{Hoosain}, M., {Blyth}, S.-L., {Skelton}, R.~E., {et~al.} 2024, \mnras, 528, 4139, \dodoi{10.1093/mnras/stae174}

\bibitem[{Horowitz(2019)}]{bootstrap}
Horowitz, J.~L. 2019, Annual Review of Economics, 11, 193, \dodoi{https://doi.org/10.1146/annurev-economics-080218-025651}

\bibitem[{{Hoyle} {et~al.}(2012){Hoyle}, {Vogeley}, \& {Pan}}]{2012MNRAS.426.3041H}
{Hoyle}, F., {Vogeley}, M.~S., \& {Pan}, D. 2012, \mnras, 426, 3041, \dodoi{10.1111/j.1365-2966.2012.21943.x}

\bibitem[{{Kauffmann} {et~al.}(2004){Kauffmann}, {White}, {Heckman}, {M{\'e}nard}, {Brinchmann}, {Charlot}, {Tremonti}, \& {Brinkmann}}]{2004MNRAS.353..713K}
{Kauffmann}, G., {White}, S. D.~M., {Heckman}, T.~M., {et~al.} 2004, \mnras, 353, 713, \dodoi{10.1111/j.1365-2966.2004.08117.x}

\bibitem[{{Kleiner} {et~al.}(2017){Kleiner}, {Pimbblet}, {Jones}, {Koribalski}, \& {Serra}}]{kleiner2017evidence}
{Kleiner}, D., {Pimbblet}, K.~A., {Jones}, D.~H., {Koribalski}, B.~S., \& {Serra}, P. 2017, \mnras, 466, 4692, \dodoi{10.1093/mnras/stw3328}

\bibitem[{{Knobel} {et~al.}(2015){Knobel}, {Lilly}, {Woo}, \& {Kova{\v{c}}}}]{2015ApJ...800...24K}
{Knobel}, C., {Lilly}, S.~J., {Woo}, J., \& {Kova{\v{c}}}, K. 2015, \apj, 800, 24, \dodoi{10.1088/0004-637X/800/1/24}

\bibitem[{{Kraljic} {et~al.}(2018{\natexlab{a}}){Kraljic}, {Arnouts}, {Pichon}, {Laigle}, {de la Torre}, {Vibert}, {Cadiou}, {Dubois}, {Treyer}, {Schimd}, {Codis}, {de Lapparent}, {Devriendt}, {Hwang}, {Le Borgne}, {Malavasi}, {Milliard}, {Musso}, {Pogosyan}, {Alpaslan}, {Bland-Hawthorn}, \& {Wright}}]{kraljic2018galaxy}
{Kraljic}, K., {Arnouts}, S., {Pichon}, C., {et~al.} 2018{\natexlab{a}}, \mnras, 474, 547, \dodoi{10.1093/mnras/stx2638}

\bibitem[{{Kraljic} {et~al.}(2018{\natexlab{b}}){Kraljic}, {Arnouts}, {Pichon}, {Laigle}, {de la Torre}, {Vibert}, {Cadiou}, {Dubois}, {Treyer}, {Schimd}, {Codis}, {de Lapparent}, {Devriendt}, {Hwang}, {Le Borgne}, {Malavasi}, {Milliard}, {Musso}, {Pogosyan}, {Alpaslan}, {Bland-Hawthorn}, \& {Wright}}]{2018MNRAS.474..547K}
---. 2018{\natexlab{b}}, \mnras, 474, 547, \dodoi{10.1093/mnras/stx2638}

\bibitem[{{Kraljic} {et~al.}(2019){Kraljic}, {Pichon}, {Dubois}, {Codis}, {Cadiou}, {Devriendt}, {Musso}, {Welker}, {Arnouts}, {Hwang}, {Laigle}, {Peirani}, {Slyz}, {Treyer}, \& {Vibert}}]{2019MNRAS.483.3227K}
{Kraljic}, K., {Pichon}, C., {Dubois}, Y., {et~al.} 2019, \mnras, 483, 3227, \dodoi{10.1093/mnras/sty3216}

\bibitem[{{Kuchner} {et~al.}(2020){Kuchner}, {Arag{\'o}n-Salamanca}, {Pearce}, {Gray}, {Rost}, {Mu}, {Welker}, {Cui}, {Haggar}, {Laigle}, {Knebe}, {Kraljic}, {Sarron}, \& {Yepes}}]{kuchner2020mapping}
{Kuchner}, U., {Arag{\'o}n-Salamanca}, A., {Pearce}, F.~R., {et~al.} 2020, \mnras, 494, 5473, \dodoi{10.1093/mnras/staa1083}

\bibitem[{{Kuchner} {et~al.}(2021){Kuchner}, {Arag{\'o}n-Salamanca}, {Rost}, {Pearce}, {Gray}, {Cui}, {Knebe}, {Rasia}, \& {Yepes}}]{kuchner2021cosmic}
{Kuchner}, U., {Arag{\'o}n-Salamanca}, A., {Rost}, A., {et~al.} 2021, \mnras, 503, 2065, \dodoi{10.1093/mnras/stab567}

\bibitem[{{Kuchner} {et~al.}(2022){Kuchner}, {Haggar}, {Arag{\'o}n-Salamanca}, {Pearce}, {Gray}, {Rost}, {Cui}, {Knebe}, \& {Yepes}}]{kuchner2022inventory}
{Kuchner}, U., {Haggar}, R., {Arag{\'o}n-Salamanca}, A., {et~al.} 2022, \mnras, 510, 581, \dodoi{10.1093/mnras/stab3419}

\bibitem[{{Kuutma} {et~al.}(2017){Kuutma}, {Tamm}, \& {Tempel}}]{kuutma2017voids}
{Kuutma}, T., {Tamm}, A., \& {Tempel}, E. 2017, \aap, 600, L6, \dodoi{10.1051/0004-6361/201730526}

\bibitem[{{Laigle} {et~al.}(2018){Laigle}, {Pichon}, {Arnouts}, {McCracken}, {Dubois}, {Devriendt}, {Slyz}, {Le Borgne}, {Benoit-L{\'e}vy}, {Hwang}, {Ilbert}, {Kraljic}, {Malavasi}, {Park}, \& {Vibert}}]{2018MNRAS.474.5437L}
{Laigle}, C., {Pichon}, C., {Arnouts}, S., {et~al.} 2018, \mnras, 474, 5437, \dodoi{10.1093/mnras/stx3055}

\bibitem[{{Lazeyras} {et~al.}(2017){Lazeyras}, {Musso}, \& {Schmidt}}]{2017JCAP...03..059L}
{Lazeyras}, T., {Musso}, M., \& {Schmidt}, F. 2017, \jcap, 2017, 059, \dodoi{10.1088/1475-7516/2017/03/059}

\bibitem[{{Li} {et~al.}(2024){Li}, {Li}, \& {Mo}}]{li2024drives}
{Li}, X., {Li}, C., \& {Mo}, H. 2024, arXiv e-prints, arXiv:2411.07977, \dodoi{10.48550/arXiv.2411.07977}

\bibitem[{{Liao} \& {Gao}(2019)}]{liao2019impact}
{Liao}, S., \& {Gao}, L. 2019, \mnras, 485, 464, \dodoi{10.1093/mnras/stz441}

\bibitem[{{Libeskind} {et~al.}(2018){Libeskind}, {van de Weygaert}, {Cautun}, {Falck}, {Tempel}, {Abel}, {Alpaslan}, {Arag{\'o}n-Calvo}, {Forero-Romero}, {Gonzalez}, {Gottl{\"o}ber}, {Hahn}, {Hellwing}, {Hoffman}, {Jones}, {Kitaura}, {Knebe}, {Manti}, {Neyrinck}, {Nuza}, {Padilla}, {Platen}, {Ramachandra}, {Robotham}, {Saar}, {Shandarin}, {Steinmetz}, {Stoica}, {Sousbie}, \& {Yepes}}]{2018MNRAS.473.1195L}
{Libeskind}, N.~I., {van de Weygaert}, R., {Cautun}, M., {et~al.} 2018, \mnras, 473, 1195, \dodoi{10.1093/mnras/stx1976}

\bibitem[{{Ma} {et~al.}(2024){Ma}, {Guo}, \& {Jones}}]{ma2024neutraluniversemachine}
{Ma}, W., {Guo}, H., \& {Jones}, M.~G. 2024, arXiv e-prints, arXiv:2409.08539, \dodoi{10.48550/arXiv.2409.08539}

\bibitem[{{Malavasi} {et~al.}(2022){Malavasi}, {Langer}, {Aghanim}, {Gal{\'a}rraga-Espinosa}, \& {Gouin}}]{2022A&A...658A.113M}
{Malavasi}, N., {Langer}, M., {Aghanim}, N., {Gal{\'a}rraga-Espinosa}, D., \& {Gouin}, C. 2022, \aap, 658, A113, \dodoi{10.1051/0004-6361/202141723}

\bibitem[{{Mandelker} {et~al.}(2021){Mandelker}, {van den Bosch}, {Springel}, {van de Voort}, {Burchett}, {Butsky}, {Nagai}, \& {Oh}}]{mandelker2021thermal}
{Mandelker}, N., {van den Bosch}, F.~C., {Springel}, V., {et~al.} 2021, \apj, 923, 115, \dodoi{10.3847/1538-4357/ac2d29}

\bibitem[{{Meng} {et~al.}(2023){Meng}, {Li}, {Mo}, {Chen}, {Jiang}, \& {Xie}}]{meng2023galaxy}
{Meng}, J., {Li}, C., {Mo}, H.~J., {et~al.} 2023, \apj, 944, 75, \dodoi{10.3847/1538-4357/acae86}

\bibitem[{{Mo} {et~al.}(2010){Mo}, {van den Bosch}, \& {White}}]{mo2010galaxy}
{Mo}, H., {van den Bosch}, F.~C., \& {White}, S. 2010, {Galaxy Formation and Evolution}

\bibitem[{{Muldrew} {et~al.}(2012){Muldrew}, {Croton}, {Skibba}, {Pearce}, {Ann}, {Baldry}, {Brough}, {Choi}, {Conselice}, {Cowan}, {Gallazzi}, {Gray}, {Gr{\"u}tzbauch}, {Li}, {Park}, {Pilipenko}, {Podgorzec}, {Robotham}, {Wilman}, {Yang}, {Zhang}, \& {Zibetti}}]{muldrew2012measures}
{Muldrew}, S.~I., {Croton}, D.~J., {Skibba}, R.~A., {et~al.} 2012, \mnras, 419, 2670, \dodoi{10.1111/j.1365-2966.2011.19922.x}

\bibitem[{{Musso} {et~al.}(2018){Musso}, {Cadiou}, {Pichon}, {Codis}, {Kraljic}, \& {Dubois}}]{2018MNRAS.476.4877M}
{Musso}, M., {Cadiou}, C., {Pichon}, C., {et~al.} 2018, \mnras, 476, 4877, \dodoi{10.1093/mnras/sty191}

\bibitem[{{Nelson} {et~al.}(2015){Nelson}, {Pillepich}, {Genel}, {Vogelsberger}, {Springel}, {Torrey}, {Rodriguez-Gomez}, {Sijacki}, {Snyder}, {Griffen}, {Marinacci}, {Blecha}, {Sales}, {Xu}, \& {Hernquist}}]{nelsondylan_2015}
{Nelson}, D., {Pillepich}, A., {Genel}, S., {et~al.} 2015, Astronomy and Computing, 13, 12, \dodoi{10.1016/j.ascom.2015.09.003}

\bibitem[{{Nelson} {et~al.}(2018){Nelson}, {Pillepich}, {Springel}, {Weinberger}, {Hernquist}, {Pakmor}, {Genel}, {Torrey}, {Vogelsberger}, {Kauffmann}, {Marinacci}, \& {Naiman}}]{nelson2018first}
{Nelson}, D., {Pillepich}, A., {Springel}, V., {et~al.} 2018, \mnras, 475, 624, \dodoi{10.1093/mnras/stx3040}

\bibitem[{{Nelson} {et~al.}(2019){Nelson}, {Springel}, {Pillepich}, {Rodriguez-Gomez}, {Torrey}, {Genel}, {Vogelsberger}, {Pakmor}, {Marinacci}, {Weinberger}, {Kelley}, {Lovell}, {Diemer}, \& {Hernquist}}]{nelson2019illustristng}
{Nelson}, D., {Springel}, V., {Pillepich}, A., {et~al.} 2019, Computational Astrophysics and Cosmology, 6, 2, \dodoi{10.1186/s40668-019-0028-x}

\bibitem[{{Nelson} {et~al.}(2020){Nelson}, {Sharma}, {Pillepich}, {Springel}, {Pakmor}, {Weinberger}, {Vogelsberger}, {Marinacci}, \& {Hernquist}}]{nelson2020resolving}
{Nelson}, D., {Sharma}, P., {Pillepich}, A., {et~al.} 2020, \mnras, 498, 2391, \dodoi{10.1093/mnras/staa2419}

\bibitem[{{O'Kane} {et~al.}(2024){O'Kane}, {Kuchner}, {Gray}, \& {Arag{\'o}n-Salamanca}}]{o2024effect}
{O'Kane}, C.~J., {Kuchner}, U., {Gray}, M.~E., \& {Arag{\'o}n-Salamanca}, A. 2024, \mnras, 534, 1682, \dodoi{10.1093/mnras/stae2142}

\bibitem[{{Park} {et~al.}(2007){Park}, {Choi}, {Vogeley}, {Gott}, {Blanton}, \& {SDSS Collaboration}}]{2007ApJ...658..898P}
{Park}, C., {Choi}, Y.-Y., {Vogeley}, M.~S., {et~al.} 2007, \apj, 658, 898, \dodoi{10.1086/511059}

\bibitem[{{Peeples} {et~al.}(2019){Peeples}, {Corlies}, {Tumlinson}, {O'Shea}, {Lehner}, {O'Meara}, {Howk}, {Earl}, {Smith}, {Wise}, \& {Hummels}}]{peeples2019figuring}
{Peeples}, M.~S., {Corlies}, L., {Tumlinson}, J., {et~al.} 2019, \apj, 873, 129, \dodoi{10.3847/1538-4357/ab0654}

\bibitem[{{Peng} {et~al.}(2010){Peng}, {Lilly}, {Kova{\v{c}}}, {Bolzonella}, {Pozzetti}, {Renzini}, {Zamorani}, {Ilbert}, {Knobel}, {Iovino}, {Maier}, {Cucciati}, {Tasca}, {Carollo}, {Silverman}, {Kampczyk}, {de Ravel}, {Sanders}, {Scoville}, {Contini}, {Mainieri}, {Scodeggio}, {Kneib}, {Le F{\`e}vre}, {Bardelli}, {Bongiorno}, {Caputi}, {Coppa}, {de la Torre}, {Franzetti}, {Garilli}, {Lamareille}, {Le Borgne}, {Le Brun}, {Mignoli}, {Perez Montero}, {Pello}, {Ricciardelli}, {Tanaka}, {Tresse}, {Vergani}, {Welikala}, {Zucca}, {Oesch}, {Abbas}, {Barnes}, {Bordoloi}, {Bottini}, {Cappi}, {Cassata}, {Cimatti}, {Fumana}, {Hasinger}, {Koekemoer}, {Leauthaud}, {Maccagni}, {Marinoni}, {McCracken}, {Memeo}, {Meneux}, {Nair}, {Porciani}, {Presotto}, \& {Scaramella}}]{2010ApJ...721..193P}
{Peng}, Y.-j., {Lilly}, S.~J., {Kova{\v{c}}}, K., {et~al.} 2010, \apj, 721, 193, \dodoi{10.1088/0004-637X/721/1/193}

\bibitem[{{Pillepich} {et~al.}(2018{\natexlab{a}}){Pillepich}, {Springel}, {Nelson}, {Genel}, {Naiman}, {Pakmor}, {Hernquist}, {Torrey}, {Vogelsberger}, {Weinberger}, \& {Marinacci}}]{2018MNRAS.473.4077P}
{Pillepich}, A., {Springel}, V., {Nelson}, D., {et~al.} 2018{\natexlab{a}}, \mnras, 473, 4077, \dodoi{10.1093/mnras/stx2656}

\bibitem[{{Pillepich} {et~al.}(2018{\natexlab{b}}){Pillepich}, {Nelson}, {Hernquist}, {Springel}, {Pakmor}, {Torrey}, {Weinberger}, {Genel}, {Naiman}, {Marinacci}, \& {Vogelsberger}}]{pillepich2018first}
{Pillepich}, A., {Nelson}, D., {Hernquist}, L., {et~al.} 2018{\natexlab{b}}, \mnras, 475, 648, \dodoi{10.1093/mnras/stx3112}

\bibitem[{{Piotrowska} {et~al.}(2022){Piotrowska}, {Bluck}, {Maiolino}, \& {Peng}}]{piotrowska2022quenching}
{Piotrowska}, J.~M., {Bluck}, A. F.~L., {Maiolino}, R., \& {Peng}, Y. 2022, \mnras, 512, 1052, \dodoi{10.1093/mnras/stab3673}

\bibitem[{{Planck Collaboration} {et~al.}(2016){Planck Collaboration}, {Ade}, {Aghanim}, {Arnaud}, {Ashdown}, {Aumont}, {Baccigalupi}, {Banday}, {Barreiro}, {Bartlett}, {Bartolo}, {Battaner}, {Battye}, {Benabed}, {Beno{\^\i}t}, {Benoit-L{\'e}vy}, {Bernard}, {Bersanelli}, {Bielewicz}, {Bock}, {Bonaldi}, {Bonavera}, {Bond}, {Borrill}, {Bouchet}, {Boulanger}, {Bucher}, {Burigana}, {Butler}, {Calabrese}, {Cardoso}, {Catalano}, {Challinor}, {Chamballu}, {Chary}, {Chiang}, {Chluba}, {Christensen}, {Church}, {Clements}, {Colombi}, {Colombo}, {Combet}, {Coulais}, {Crill}, {Curto}, {Cuttaia}, {Danese}, {Davies}, {Davis}, {de Bernardis}, {de Rosa}, {de Zotti}, {Delabrouille}, {D{\'e}sert}, {Di Valentino}, {Dickinson}, {Diego}, {Dolag}, {Dole}, {Donzelli}, {Dor{\'e}}, {Douspis}, {Ducout}, {Dunkley}, {Dupac}, {Efstathiou}, {Elsner}, {En{\ss}lin}, {Eriksen}, {Farhang}, {Fergusson}, {Finelli}, {Forni}, {Frailis}, {Fraisse}, {Franceschi}, {Frejsel}, {Galeotta}, {Galli}, {Ganga}, {Gauthier}, {Gerbino}, {Ghosh}, {Giard},
  {Giraud-H{\'e}raud}, {Giusarma}, {Gjerl{\o}w}, {Gonz{\'a}lez-Nuevo}, {G{\'o}rski}, {Gratton}, {Gregorio}, {Gruppuso}, {Gudmundsson}, {Hamann}, {Hansen}, {Hanson}, {Harrison}, {Helou}, {Henrot-Versill{\'e}}, {Hern{\'a}ndez-Monteagudo}, {Herranz}, {Hildebrandt}, {Hivon}, {Hobson}, {Holmes}, {Hornstrup}, {Hovest}, {Huang}, {Huffenberger}, {Hurier}, {Jaffe}, {Jaffe}, {Jones}, {Juvela}, {Keih{\"a}nen}, {Keskitalo}, {Kisner}, {Kneissl}, {Knoche}, {Knox}, {Kunz}, {Kurki-Suonio}, {Lagache}, {L{\"a}hteenm{\"a}ki}, {Lamarre}, {Lasenby}, {Lattanzi}, {Lawrence}, {Leahy}, {Leonardi}, {Lesgourgues}, {Levrier}, {Lewis}, {Liguori}, {Lilje}, {Linden-V{\o}rnle}, {L{\'o}pez-Caniego}, {Lubin}, {Mac{\'\i}as-P{\'e}rez}, {Maggio}, {Maino}, {Mandolesi}, {Mangilli}, {Marchini}, {Maris}, {Martin}, {Martinelli}, {Mart{\'\i}nez-Gonz{\'a}lez}, {Masi}, {Matarrese}, {McGehee}, {Meinhold}, {Melchiorri}, {Melin}, {Mendes}, {Mennella}, {Migliaccio}, {Millea}, {Mitra}, {Miville-Desch{\^e}nes}, {Moneti}, {Montier}, {Morgante}, {Mortlock},
  {Moss}, {Munshi}, {Murphy}, {Naselsky}, {Nati}, {Natoli}, {Netterfield}, {N{\o}rgaard-Nielsen}, {Noviello}, {Novikov}, {Novikov}, {Oxborrow}, {Paci}, {Pagano}, {Pajot}, {Paladini}, {Paoletti}, {Partridge}, {Pasian}, {Patanchon}, {Pearson}, {Perdereau}, {Perotto}, {Perrotta}, {Pettorino}, {Piacentini}, {Piat}, {Pierpaoli}, {Pietrobon}, {Plaszczynski}, {Pointecouteau}, {Polenta}, {Popa}, {Pratt}, \& {Pr{\'e}zeau}}]{ade2016planck}
{Planck Collaboration}, {Ade}, P.~A.~R., {Aghanim}, N., {et~al.} 2016, \aap, 594, A13, \dodoi{10.1051/0004-6361/201525830}

\bibitem[{{Popesso} {et~al.}(2011){Popesso}, {Rodighiero}, {Saintonge}, {Santini}, {Grazian}, {Lutz}, {Brusa}, {Altieri}, {Andreani}, {Aussel}, {Berta}, {Bongiovanni}, {Cava}, {Cepa}, {Cimatti}, {Daddi}, {Dominguez}, {Elbaz}, {F{\"o}rster Schreiber}, {Genzel}, {Gruppioni}, {Magdis}, {Maiolino}, {Magnelli}, {Nordon}, {P{\'e}rez Garc{\'\i}a}, {Poglitsch}, {Pozzi}, {Riguccini}, {Sanchez-Portal}, {Shao}, {Sturm}, {Tacconi}, {Valtchanov}, {Wieprecht}, \& {Wetzstein}}]{popesso2011effect}
{Popesso}, P., {Rodighiero}, G., {Saintonge}, A., {et~al.} 2011, \aap, 532, A145, \dodoi{10.1051/0004-6361/201015672}

\bibitem[{{Primack}(2024)}]{primack2024galaxy}
{Primack}, J.~R. 2024, Annual Review of Nuclear and Particle Science, 74, 173, \dodoi{10.1146/annurev-nucl-102622-023052}

\bibitem[{{Rojas} {et~al.}(2004){Rojas}, {Vogeley}, {Hoyle}, \& {Brinkmann}}]{2004ApJ...617...50R}
{Rojas}, R.~R., {Vogeley}, M.~S., {Hoyle}, F., \& {Brinkmann}, J. 2004, \apj, 617, 50, \dodoi{10.1086/425225}

\bibitem[{{Schaye} {et~al.}(2015){Schaye}, {Crain}, {Bower}, {Furlong}, {Schaller}, {Theuns}, {Dalla Vecchia}, {Frenk}, {McCarthy}, {Helly}, {Jenkins}, {Rosas-Guevara}, {White}, {Baes}, {Booth}, {Camps}, {Navarro}, {Qu}, {Rahmati}, {Sawala}, {Thomas}, \& {Trayford}}]{schaye2015eagle}
{Schaye}, J., {Crain}, R.~A., {Bower}, R.~G., {et~al.} 2015, \mnras, 446, 521, \dodoi{10.1093/mnras/stu2058}

\bibitem[{{Sobral} {et~al.}(2011){Sobral}, {Best}, {Smail}, {Geach}, {Cirasuolo}, {Garn}, \& {Dalton}}]{sobral2011dependence}
{Sobral}, D., {Best}, P.~N., {Smail}, I., {et~al.} 2011, \mnras, 411, 675, \dodoi{10.1111/j.1365-2966.2010.17707.x}

\bibitem[{{Song} {et~al.}(2021){Song}, {Laigle}, {Hwang}, {Devriendt}, {Dubois}, {Kraljic}, {Pichon}, {Slyz}, \& {Smith}}]{song2021beyond}
{Song}, H., {Laigle}, C., {Hwang}, H.~S., {et~al.} 2021, \mnras, 501, 4635, \dodoi{10.1093/mnras/staa3981}

\bibitem[{{Sousbie}(2011)}]{sousbie2011persistent}
{Sousbie}, T. 2011, \mnras, 414, 350, \dodoi{10.1111/j.1365-2966.2011.18394.x}

\bibitem[{{Sousbie} {et~al.}(2011){Sousbie}, {Pichon}, \& {Kawahara}}]{sousbieet2011persistent}
{Sousbie}, T., {Pichon}, C., \& {Kawahara}, H. 2011, \mnras, 414, 384, \dodoi{10.1111/j.1365-2966.2011.18395.x}

\bibitem[{{Springel}(2010)}]{springel2010pur}
{Springel}, V. 2010, \mnras, 401, 791, \dodoi{10.1111/j.1365-2966.2009.15715.x}

\bibitem[{{Springel} {et~al.}(2018){Springel}, {Pakmor}, {Pillepich}, {Weinberger}, {Nelson}, {Hernquist}, {Vogelsberger}, {Genel}, {Torrey}, {Marinacci}, \& {Naiman}}]{springel2018first}
{Springel}, V., {Pakmor}, R., {Pillepich}, A., {et~al.} 2018, \mnras, 475, 676, \dodoi{10.1093/mnras/stx3304}

\bibitem[{{Stevens} {et~al.}(2019){Stevens}, {Diemer}, {Lagos}, {Nelson}, {Obreschkow}, {Wang}, \& {Marinacci}}]{stevens2019origin}
{Stevens}, A. R.~H., {Diemer}, B., {Lagos}, C. d.~P., {et~al.} 2019, \mnras, 490, 96, \dodoi{10.1093/mnras/stz2513}

\bibitem[{{Tempel} {et~al.}(2014{\natexlab{a}}){Tempel}, {Stoica}, {Mart{\'\i}nez}, {Liivam{\"a}gi}, {Castellan}, \& {Saar}}]{2014MNRAS.438.3465T}
{Tempel}, E., {Stoica}, R.~S., {Mart{\'\i}nez}, V.~J., {et~al.} 2014{\natexlab{a}}, \mnras, 438, 3465, \dodoi{10.1093/mnras/stt2454}

\bibitem[{{Tempel} {et~al.}(2014{\natexlab{b}}){Tempel}, {Stoica}, {Mart{\'\i}nez}, {Liivam{\"a}gi}, {Castellan}, \& {Saar}}]{tempel2014detecting}
---. 2014{\natexlab{b}}, \mnras, 438, 3465, \dodoi{10.1093/mnras/stt2454}

\bibitem[{{Thomas} {et~al.}(2005){Thomas}, {Maraston}, {Bender}, \& {Mendes de Oliveira}}]{thomas2005epochs}
{Thomas}, D., {Maraston}, C., {Bender}, R., \& {Mendes de Oliveira}, C. 2005, \apj, 621, 673, \dodoi{10.1086/426932}

\bibitem[{{Torrey} {et~al.}(2019){Torrey}, {Vogelsberger}, {Marinacci}, {Pakmor}, {Springel}, {Nelson}, {Naiman}, {Pillepich}, {Genel}, {Weinberger}, \& {Hernquist}}]{torrey2019evolution}
{Torrey}, P., {Vogelsberger}, M., {Marinacci}, F., {et~al.} 2019, \mnras, 484, 5587, \dodoi{10.1093/mnras/stz243}

\bibitem[{{van de Weygaert} \& {Bond}(2008)}]{2008LNP...740..335V}
{van de Weygaert}, R., \& {Bond}, J.~R. 2008, in A Pan-Chromatic View of Clusters of Galaxies and the Large-Scale Structure, ed. M.~{Plionis}, O.~{L{\'o}pez-Cruz}, \& D.~{Hughes}, Vol. 740, 335, \dodoi{10.1007/978-1-4020-6941-3_10}

\bibitem[{{Vulcani} {et~al.}(2019){Vulcani}, {Poggianti}, {Moretti}, {Gullieuszik}, {Fritz}, {Franchetto}, {Fasano}, {Bettoni}, \& {Jaff{\'e}}}]{vulcani2019gasp}
{Vulcani}, B., {Poggianti}, B.~M., {Moretti}, A., {et~al.} 2019, \mnras, 487, 2278, \dodoi{10.1093/mnras/stz1399}

\bibitem[{{Wang} {et~al.}(2023){Wang}, {Wang}, \& {Chen}}]{wang2023environmental}
{Wang}, K., {Wang}, X., \& {Chen}, Y. 2023, \apj, 951, 66, \dodoi{10.3847/1538-4357/acd633}

\bibitem[{{Wechsler} \& {Tinker}(2018)}]{2018ARA&A..56..435W}
{Wechsler}, R.~H., \& {Tinker}, J.~L. 2018, \araa, 56, 435, \dodoi{10.1146/annurev-astro-081817-051756}

\bibitem[{{White} \& {Frenk}(1991)}]{1991ApJ...379...52W}
{White}, S. D.~M., \& {Frenk}, C.~S. 1991, \apj, 379, 52, \dodoi{10.1086/170483}

\bibitem[{{White} \& {Rees}(1978)}]{white1978core}
{White}, S.~D.~M., \& {Rees}, M.~J. 1978, \mnras, 183, 341, \dodoi{10.1093/mnras/183.3.341}

\bibitem[{{Winkel} {et~al.}(2021){Winkel}, {Pasquali}, {Kraljic}, {Smith}, {Gallazzi}, \& {Jackson}}]{winkel2021imprint}
{Winkel}, N., {Pasquali}, A., {Kraljic}, K., {et~al.} 2021, \mnras, 505, 4920, \dodoi{10.1093/mnras/stab1562}

\bibitem[{{Wolfire} {et~al.}(2003){Wolfire}, {McKee}, {Hollenbach}, \& {Tielens}}]{2003ApJ...587..278W}
{Wolfire}, M.~G., {McKee}, C.~F., {Hollenbach}, D., \& {Tielens}, A.~G.~G.~M. 2003, \apj, 587, 278, \dodoi{10.1086/368016}

\bibitem[{{Xie} {et~al.}(2020){Xie}, {De Lucia}, {Hirschmann}, \& {Fontanot}}]{xie2020influence}
{Xie}, L., {De Lucia}, G., {Hirschmann}, M., \& {Fontanot}, F. 2020, \mnras, 498, 4327, \dodoi{10.1093/mnras/staa2370}

\bibitem[{{Zakharova} {et~al.}(2024){Zakharova}, {Vulcani}, {De Lucia}, {Finn}, {Rudnick}, {Combes}, {Castignani}, {Fontanot}, {Jablonka}, {Xie}, \& {Hirschmann}}]{zakharova2024virgo}
{Zakharova}, D., {Vulcani}, B., {De Lucia}, G., {et~al.} 2024, \aap, 690, A300, \dodoi{10.1051/0004-6361/202450825}

\bibitem[{{Zel'dovich}(1970)}]{1970A&A.....5...84Z}
{Zel'dovich}, Y.~B. 1970, \aap, 5, 84

\bibitem[{{Zheng} {et~al.}(2022){Zheng}, {Liao}, {Hu}, {Gao}, {Grand}, {Gu}, \& {Guo}}]{2022MNRAS.514.2488Z}
{Zheng}, H., {Liao}, S., {Hu}, J., {et~al.} 2022, \mnras, 514, 2488, \dodoi{10.1093/mnras/stac1476}

\bibitem[{{Zhu} \& {Feng}(2017)}]{2017ApJ...838...21Z}
{Zhu}, W., \& {Feng}, L.-L. 2017, \apj, 838, 21, \dodoi{10.3847/1538-4357/aa61f9}

\bibitem[{{Zhu} {et~al.}(2024){Zhu}, {Wang}, {Zhang}, {Zheng}, \& {Feng}}]{2024ApJ...967..141Z}
{Zhu}, W., {Wang}, T.-R., {Zhang}, F., {Zheng}, Y., \& {Feng}, L.-L. 2024, \apj, 967, 141, \dodoi{10.3847/1538-4357/ad3e6a}

\bibitem[{{Zhu} {et~al.}(2021){Zhu}, {Zhang}, \& {Feng}}]{2021ApJ...920....2Z}
{Zhu}, W., {Zhang}, F., \& {Feng}, L.-L. 2021, \apj, 920, 2, \dodoi{10.3847/1538-4357/ac15f1}

\bibitem[{{Zhu} {et~al.}(2022){Zhu}, {Zhang}, \& {Feng}}]{2022ApJ...924..132Z}
---. 2022, \apj, 924, 132, \dodoi{10.3847/1538-4357/ac37b9}

\bibitem[{{Zinger} {et~al.}(2020){Zinger}, {Pillepich}, {Nelson}, {Weinberger}, {Pakmor}, {Springel}, {Hernquist}, {Marinacci}, \& {Vogelsberger}}]{zinger2020ejective}
{Zinger}, E., {Pillepich}, A., {Nelson}, D., {et~al.} 2020, \mnras, 499, 768, \dodoi{10.1093/mnras/staa2607}

\end{thebibliography}
\bibliographystyle{aasjournal}

\appendix 
\section{filament segments}

The filaments identified by \texttt{DisPerSe} using our procedure range in length from approximately $1\,\mathrm{Mpc}/h$ to over $10\,\mathrm{Mpc}/h$. \texttt{DisPerSe} generates continuous segment points along each filament that connecting two nodes. To divide the filaments into segments of length of $1.5\,\mathrm{Mpc}/h$ to $2.5\,\mathrm{Mpc}/h$, we follow the procedure outlined below:

\begin{enumerate}
    \item For each filament, we begin at one of its two nodes and sequentially traverse the segment points in order. At each segment point, we compute the angle, $\theta_{adj}$, ormed between two consecutive short segments identified by \texttt{DisPerSe}. Additionally, we track the cumulative length from the starting node.
    
    \item For significantly curved filaments,those with at least one adjacent segment angle $\theta_{adj} \geq 45^\circ$, we split the filaments at segment points where $\theta_{adj} \geq 45^\circ$). After splitting, we retain segments with lengths between  $1.5\,\mathrm{Mpc}/h$ and $2.5\,\mathrm{Mpc}/h$.  For segments longer than $2.5\,\mathrm{Mpc}/h$, we further divide it into pieces with length between $1.5\,\mathrm{Mpc}/h$ and $2.5\,\mathrm{Mpc}/h$. Segments shorter than $1.5\,\mathrm{Mpc}/h$ are discarded, as they are typically less massive.
    
    \item For relatively straight filaments, those where the adjacent segment angles at various segment points, as well as the averaged segment angles, remain consistently below $45^\circ$ along the entire filament, we break filaments according to the cumulative length. we divide them based on cumulative length. Specifically, we break the filament at selected segment points to ensure that the resulting segments have lengths between $1.5\,\mathrm{Mpc}/h$ and $2.5\,\mathrm{Mpc}/h$. The left segments shorter than $1.5\,\mathrm{Mpc}/h$ are also discarded.  

\end{enumerate}

After processing, we generated a sample of filament segments, each with a length within the range of $1.5$ to $2.5\,\mathrm{Mpc}/h$. Each segment in our sample represents a combination of multiple the initially identified DisPerSe segments. For each filament segment, we define the spine as the straight line connecting its two endpoints. Approximating the segment as a cylinder, we define the upper and lower surfaces as planes perpendicular to the spine, passing through the endpoints. To determine the segment's thickness (cylinder radius), we calculate the radial gas density profile. Specifically, using the spine as the central axis, we compute the gas mass within cylindrical shells and normalize by their respective volumes. We apply the same procedure to obtain the dark matter density profiles.

We determine the thickness of filament segments using their radial dark matter density profiles. Specifically, the filament thickness, denoted as $R_{\mathrm{fil}}$, is defined as the radius where the dark matter density reaches the cosmic mean. We then calculate the total mass of each filament segment by summing the dark matter and gas particles within this cylindrical volume. In our analysis, we emphasize the influence of a galaxy's nearest massive filament segment when evaluating the effects of cosmic filaments.

\section{Miscellaneous}

\begin{figure}[h]
\begin{centering}
\includegraphics[width=0.75\textwidth]{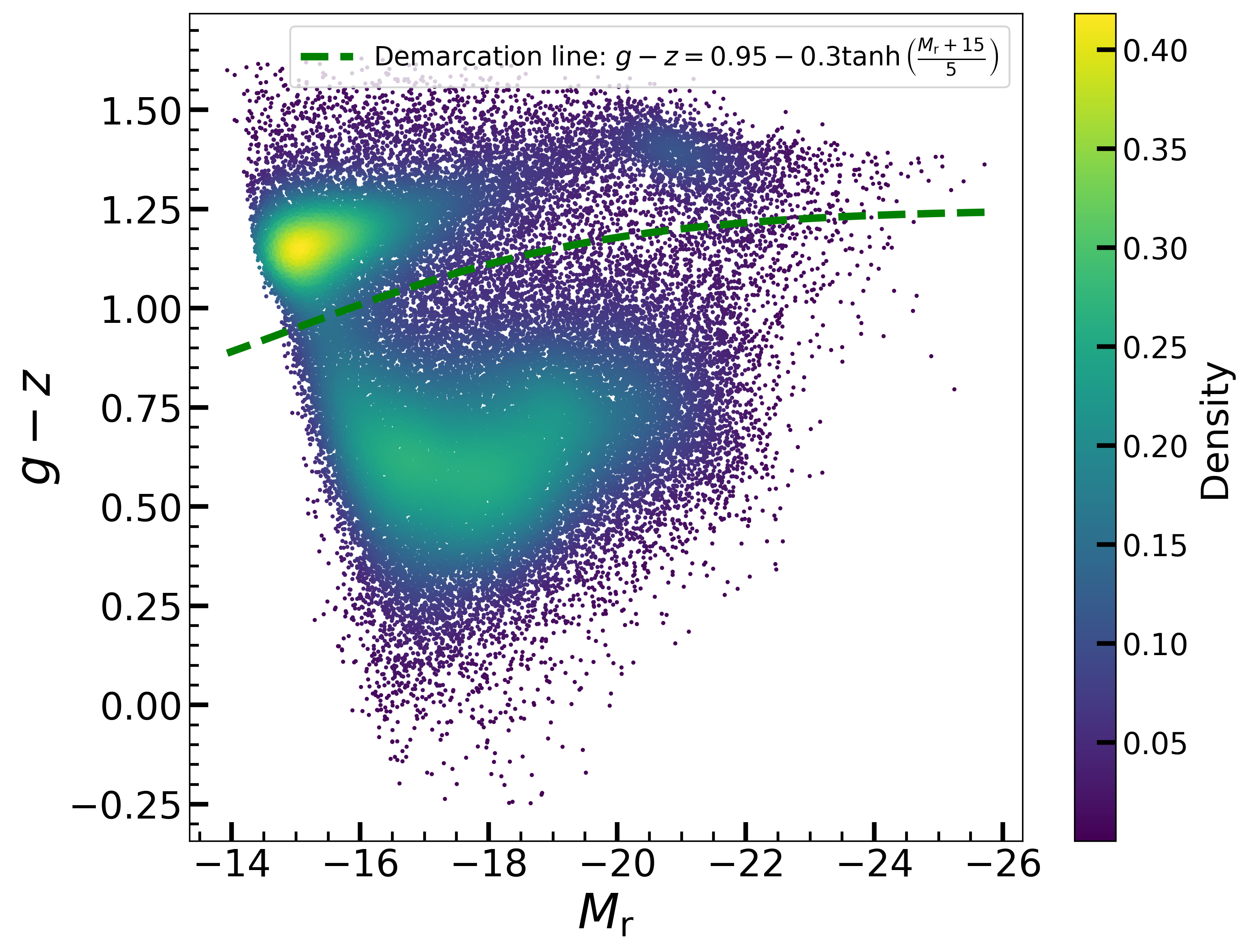}
\caption{Kernel density estimation (KDE) distribution of galaxies in the $ (g-z) - M_\mathrm{r}$ parameter space. The demarcation line separates red and blue galaxy sequences.}
\label{fig:color_TNG100}
\end{centering}
\end{figure}

\begin{figure}[h]
\begin{centering}
\includegraphics[width=0.75\textwidth]{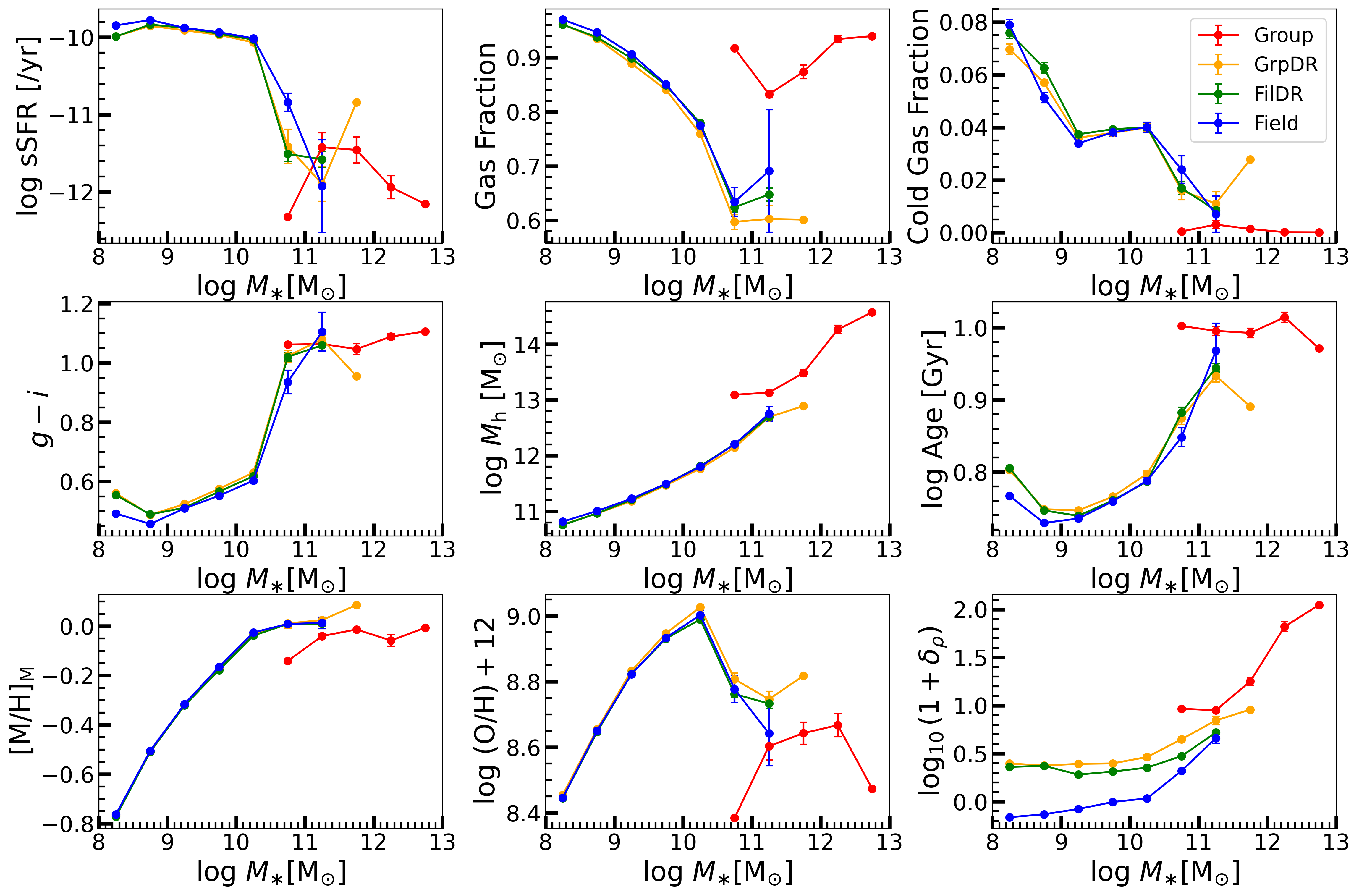}
\caption{Same as Figure \ref{fig:porperties_central}, but for cental galaxies with $D_\mathrm{grp} > 2R_\mathrm{h}$.}
\label{fig:property_StellarMass_central_TNG100_out2R}
\end{centering}
\end{figure}

\begin{figure}[h]
\begin{centering}
\includegraphics[width=0.75\textwidth]{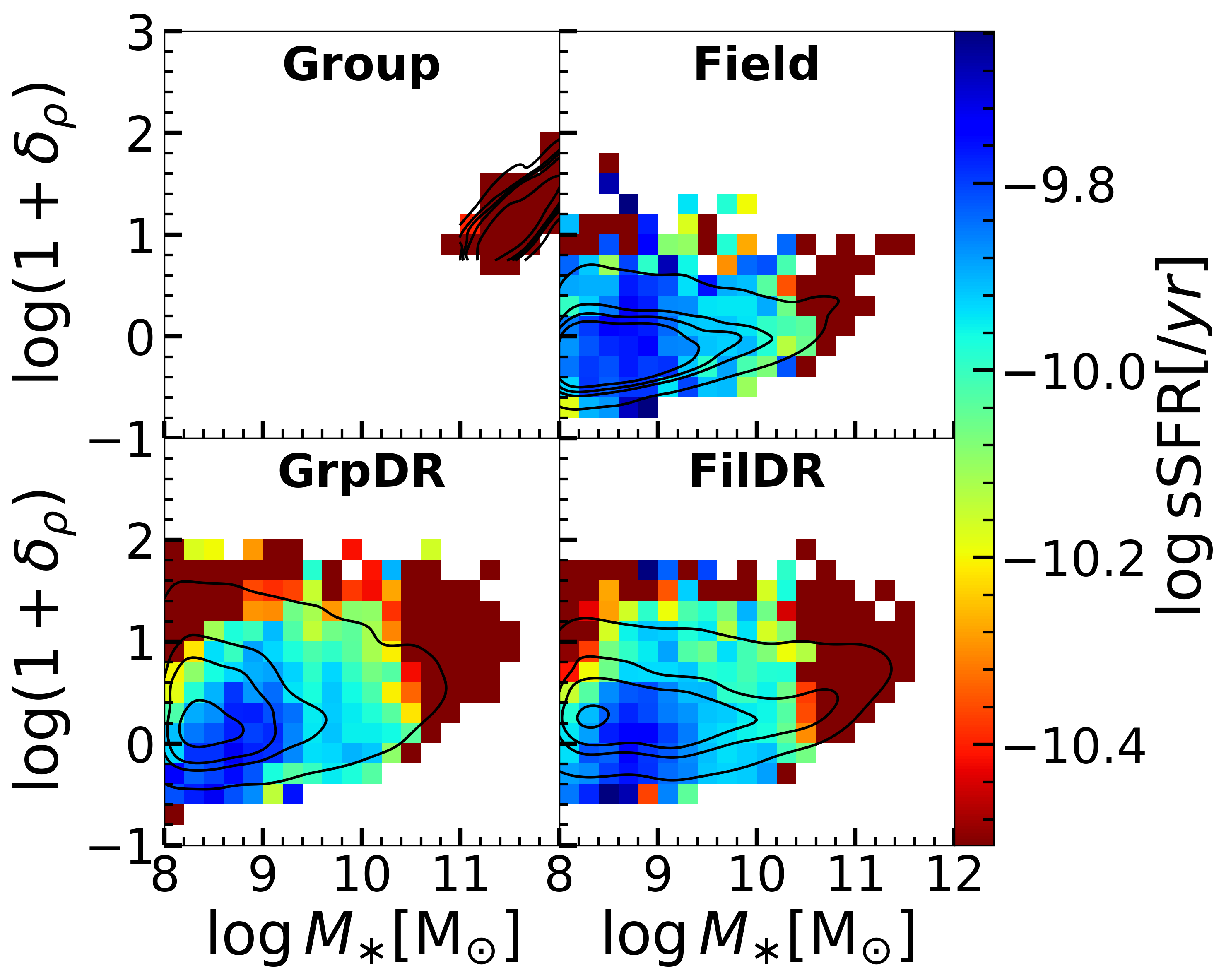}
\caption{Same as Figure \ref{fig:overdensity_correction_central}, but for cental galaxies with $D_\mathrm{grp} > 2R_\mathrm{h}$.}
\label{fig:distribution_density_mass_out2R}
\end{centering}
\end{figure}





\end{document}